\documentclass{jfm}
\usepackage{amsmath}
\usepackage{amsfonts}
\usepackage{amssymb}
\usepackage[utf8x]{inputenc}
\usepackage{graphicx}
\usepackage{natbib}
\usepackage{float}
\usepackage{comment}
\usepackage{epstopdf}
\usepackage{lineno}
\usepackage{scalerel,stackengine}
\usepackage{color}

\title{Phase transition of the energy flux in the near-inertial wave--mesoscale eddy coupled turbulence}
\author{Jin-Han Xie $^1$ }
\affiliation{$^1$ Department of Mechanics and Engineering Science at College of Engineering, \\
	State Key Laboratory for Turbulence and Complex Systems\\ and Beijing Innovation Center for Engineering Science and Advanced Technology,\\ Peking University, Beijing, 100871, PR China}

\newcommand{\av}[1]{\left<#1\right>}

\newcommand{\br}[1]{\left( #1 \right)}
\newcommand{\sbr}[1]{\left[ #1 \right]}
\newcommand{\lbr}[1]{\left\{ #1 \right\}}
\newcommand{\ex}{\mathrm{e}}
\newcommand{\ii}{\mathrm{i}}

\newcommand{\Bu}{\textrm{Bu}}

\newcommand{\mc}[1]{\mathcal{#1}}
\newcommand{\abs}[1]{\left| #1 \right|}

\newcommand{\bx}{\boldsymbol{x}}

\newcommand{\bk}{\boldsymbol{k}}

\newcommand{\pa}[1]{\partial_{#1}}
\newcommand{\Ja}[2]{\mathcal{J}\left(#1,#2\right)}

\newcommand{\dd}{\mathrm{d}}

\begin{document}
	\maketitle
	\begin{abstract}
		Wind forcing injects energy into the mesoscale eddies and near-inertial waves (NIWs) in the ocean, and the NIW is believed to solve the puzzle of mesoscale energy budget by absorbing energy from mesoscale eddies followed by a forward cascade of NIW energy which finally dissipates at the ocean interior.
		This work studies the turbulent energy transfer in the NIW--quasigeostrophic mean mesoscale eddy coupled system based on a previously derived two-dimensional model which has a Hamiltonian structure and inherits conserved quantities in the Boussinesq equations (Xie \& Vanneste, \textit{J. Fluid Mech.}, vol. 774, 2015, pp. 147--169). 
		Based on the conservation of energy, potential enstrophy and wave action, we propose a heuristic argument predicting the existence of phase transition with changing the relative strength between NIW and mean flow.
		By running forced-dissipative numerical simulations with varying parameter $R$, the ratio of the magnitude of NIW and mean-flow forcing, we justify the existence of phase transition, which is found to be second-order, around critical value $R_c$.
		When $0<R<R_c$, energy transfers bidirectionally, wave action transfers downscale, and vorticity form strong cyclones.
		While when $R>R_c$, energy transfers downscale, wave action transfers bidirectionally, and vortex filaments are dominant.
		We find the catalytic wave induction (CWI) mechanism where the NIW induces a downscale energy flux of the mean flow.
		The CWI mechanism differs from the stimulated loss of balance by the absence of energy conversion from the mesoscale eddy to NIW, and it is found to be effective in the toy-model study, making it potentially important for ocean energetics.
		
	\end{abstract}	
	
	\section{Introduction}

	The mesoscale eddy, which has horizontal scale from one to hundreds of kilometres, contains a significant part of the ocean energy, however, its energy budget remains not well understood: wind forcing and large-scale circulations injection energy into the mesoscale eddies, and the mesoscale eddies dissipate at the boundaries and be converted to other types of motions, but the known amount of energy injection is much larger than that of energy dissipation \citep{Wuns2004,Ferr2009}. Several candidates are thought to be responsible for the mesoscale eddy energy closure, such as the simultaneous loss of balance \cite[cf.][]{Vanneste2013} and the stimulated near-inertial wave (NIW) generation, which is also named stimulated loss of balance \citep{Xie2015}.
	
	There are a variety of mechanisms potentially explaining that NIWs can extract energy from balanced flow, and they have been studied in various setups and parameter regimes. 
	\cite{Gertz2009} run numerical simulations in a periodic box with O(1) Rossby number and find that the energy transfer between 2D and 3D motions is crucial.
	Also in the parameter regime with O(1) Rossby number, \cite{Taylor2016} consider a channel flow with external forcing acting on both low and high frequencies, and they find that the Reynolds stresses of NIW act as a kinetic energy sink for mesoscale motions. 
	In \cite{Barkan2017}, numerical simulations in a channel flow with external forcing find the direct extraction dominates while the mechanism of stimulated NIW generation is subdominant.
	NIW is also found to absorb mean flow energy in frontogenesis \citep{Thomas2012}.
	
	Applying the generalised-lagrangian mean theory \citep{Andrew1978,Soward2010} in a variational framework \citep{Salmon1988,Salmon2013,Salmon2016}, \cite{Xie2015} derive an asymptotic coupled model in the parameter regime of small Rossby and Burger number to describe the interaction between NIWs and quasigeostrophic (QG) mean flow. 
	This model naturally couples the classic models of QG mean flow \cite[cf.][]{Salmon1998b} and the Young-Ben Jelloul (YBJ) equation for the slow evolution of NIW magnitude \citep{Young1997}, which have been validated in decades of study.
	It has been extended to include the second harmonic of inertial oscillations by \cite{Wagner2016} using the concept of available potential vorticity \citep{Wagner2015} and been used to study the details of the stimulated loss of balance \citep{Rocha2018}.
	
	The variational framework ensures the model's Hamiltonian structure, and therefore it preserves conservation laws of the original Boussinesq system by inheriting the symmetries, making this model to capture certain key mechanisms in the ocean energetics and to study statistics. 
	So in this paper, we make use of the two-dimensional (2D) version of this model, where the QG flow is barotropic and the NIW has only one single vertical wavenumber, to study the direction of energy transfer with changing ratio of external energy injection into the NIW and QG components. 
	And we run forced-dissipative numerical simulations to statistically steady states such that the energy injection balances the energy dissipation.
	
	The direction of energy transfer across scales is a basic question to turbulence.
	A key feature of the atmospheric and oceanic turbulence is the bidirectional energy transfer, which has been observed in numerical simulations of the rotating stratified turbulence \cite[e.g.][]{Marino2015,Pouquet2017,Alexakis2018}. 
	Even though the turbulence theory linking bidirectional energy transfer with measurable structure functions has been proposed \citep{XieBuhler2019b}, the mechanism for bidirectional energy transfer is less understood. 
	The two primary reasons for bidirectional energy transfer are the impact of inertia--gravity waves and the short aspect ratio \citep{Benavides2017}. 
	The toy model of NIW-QG interaction studied in this paper provides a simple setup to study the wave-induced bidirectional energy transfer. 
	
	The paper is organised as follows. In \S \ref{sec_arguement} we provide a heuristic argument based on the preserved quantities to predict the transfer direction of preserved quantities in the NIW-QG model derived by \citet{Xie2015}. 
	We show the results of forced-dissipative numerical simulations of YBJ$^+$-QG model, which avoids the ``ultraviolet catastrophe" of the NIW-QG model, in \S \ref{sec_num} to confirm the existence of bidirectional energy and wave action transfer.
	We discuss and summarize our results in \S \ref{sec_dis} and $\S$ \ref{sec_sum}, respectively. 
	Appendix \ref{sec_YBJ+} provides a derivation of the YBJ$^+$-QG model in a variational approach and discuss the directions of conserved quantities in this model.
	
	\section{A heuristic argument for the direction of energy transfer} \label{sec_arguement}
	
	In this section, we propose a heuristic argument applying to a forced-dissipative turbulent system with three inviscid preserved quantities to predict the transfer direction of preserved quantities across scales.
	
	
	We focus on the dynamics of the interaction between NIW and QG mean flow using the model derived by \citet{Xie2015}. 
	Since the NIW is described by the slow modulation of the wave amplitude with a fixed frequency equaling to the Coriolis frequency, the model also preserves wave action due to the phase invariant. 
	Even though the wave action is only asymptotically conserved in the original system, we believe it is still crucial for the dynamics of the wave-mean flow coupled system considering that the NIW peak is notable in the ocean observation \cite[cf.][]{Ferr2009}.
	
	In this paper, for simplicity and numerical efficiency, we focus on the 2D version of the YBJ-QG model on an $f$-plane, where the QG mean flow is barotropic ($z$-independent), and the NIW has a single mode in vertical. 
	The governing equations are \cite[cf.][]{Xie2015} 
	\begin{subequations}\label{NIWQG0}
		\begin{align}
		M_t + \Ja{\psi}{M} - \frac{\ii N^2}{2m^2 f}\nabla^2M + \frac{\ii}{2}M\nabla^2\psi &= 0,\label{YBJ0}\\
		q_t + \Ja{\psi}{q} & = 0,\\
		\mathrm{with}\quad q = \nabla^2 \psi + \frac{\ii f}{2}\Ja{M^*}{M} + \frac{f}{4}\nabla^2 |M|^2,
		\end{align}
	\end{subequations}
	where $M$ is the complex wave amplitude such that $u+\ii v = -\ii f M\ex^{-\ii f t+\ii m z}$ with $u$ and $v$ the two horizontal velocities, $\mc{J}$ denotes the Jacobian, $N$ is the Brunt-V\"ais\"al\"a (buoyancy) frequency, $m$ is the vertical wavenumber of NIW, $f$ is the Coriolis frequency, $\psi$ is the stream function of QG mean flow.
	
	The 2D model (\ref{NIWQG0}) preserves the energy, potential enstrophy and wave action:
	\begin{subequations}\label{Conservation0}
		\begin{align}
		\mc{E} &= \int \lbr{\frac{1}{2}|\nabla\psi|^2+\frac{1}{4}\frac{N^2}{m^2}|\nabla M|^2} \dd \bx,\\
		\mc{P} &= \int  q^2 \dd \bx,\\
		\mc{A} &= \int |M|^2\dd \bx.
		\end{align}
	\end{subequations}
	The total energy $\mc{E}$ contains two quadratic forms of mean flow and wave, and we name them $\mc{E}_{QG}$ and $\mc{E}_{W}$, respectively. 
	They are defined as 
	\begin{equation}\label{energy_decomp}
	\mc{E}_{QG} = \int \frac{1}{2}|\nabla\psi|^2 \dd \bx \quad \mathrm{and} \quad 
	\mc{E}_{W} = \int \frac{1}{4}\frac{N^2}{m^2}|\nabla M|^2 \dd \bx,
	\end{equation}
	therefore $\mc{E}=\mc{E}_{QG}+\mc{E}_{W}$.
	The potential enstrophy $\mc{P}$ is the integration of the quadratic of PV therefore it contains quadratic, cubic and quartic terms of both mean flow and wave amplitude.
	But the wave action $\mc{A}$ is purely a wave quadratic quantity.
	
	To understand the transfer direction of conserved quantities, we consider a turbulent picture similar to those proposed by \citet{Kraichnan1967} and \citet{Eyink1996} to show the direction of energy and enstrophy cascades in 2D turbulence, and the one used in wave turbulence \cite[cf. Chapter 1 in][]{Cardy2008}. 
	However, different from 2D turbulence and wave turbulence, our system (\ref{NIWQG0}) preserves three quantities. We consider that the energy, wave action and potential enstrophy are injected into the system at an intermediate wavenumber $k_f$ and they dissipate at both a small wavenumber $k_1$ and a large wavenumber $k_2$. 
	We denote the energy, potential enstrophy and wave action dissipations at these two wavenumbers as $\mc{E}_{i}$, $\mc{P}_{i}$ and $\mc{A}_i$ ($i=1,2$), respectively.  
	Therefore, the three conversations (\ref{Conservation0}) imply that
	\begin{subequations}\label{balance}
		\begin{align}
		\mc{E}_f = \mc{E}_1 + \mc{E}_2, \quad
		\mc{P}_f = \mc{P}_1 + \mc{P}_2, \quad \mathrm{and} \quad
		\mc{A}_f = \mc{A}_1 + \mc{A}_2,
		\end{align}
	\end{subequations}
	where the symbols with lower index ``$f$" denote the injections at the forcing wavenumber.
	
	However, different from the 2D turbulence case, where due to the quadratic energy and enstrophy the transfer direction of energy and enstrophy are determined by solving (\ref{balance}), we cannot similarly determine the transfer directions of these three quantities, which is the main topic of this paper, and we will show that the preservation of three quantities brings about a more complicated picture for the transfer directions.

	We first consider two limiting cases that hint us the transfer directions in general cases. 
	The first case is the mean-flow-dominant case where the NIW is week compared with the mean flow, therefore the wave action as a purely wave effect is not dominant, also, the wave effect is of high order in the energy and potential enstrophy. 
	Then, to the leading order, the mean flow dominates the energy and potential enstrophy
	\begin{subequations}
		\begin{align}
		\mc{E} &= \int \lbr{\frac{1}{2}|\nabla\psi|^2} \dd \bx,\\
		\mc{P} &= \int  |\nabla^2\psi|^2 \dd \bx.
		\end{align}
	\end{subequations}
	Therefore the system recovers the scenario of 2D turbulence: energy transfers upscale and potential enstrophy transfers downscale. Also, because the feedback to the mean flow is weak, NIW behaves like a passive scalar and the wave action transfers downscale.
	
	The other extreme is the NIW-dominant case, where the mean flow is weak compared with the NIW. In this case, the potential enstrophy is dominated by the wave quartic terms and is assumed to be subdominant, therefore, similar to the wave turbulence, energy and wave action control the turbulent dynamics. 
	To the leading order energy and wave action are expressed as
	\begin{subequations}\label{conserved_quantities}
		\begin{align}
		\mc{E} &= \frac{1}{4}\frac{N^2}{m^2} \int |\nabla M|^2 \dd \bx, \\
		\mc{A} &= \int |M|^2\dd \bx.
		\end{align}
	\end{subequations}
	Therefore, the conservations (\ref{balance}) imply that
	\begin{subequations}\label{balance1}
		\begin{align}
		\mc{E}_f &= \frac{1}{4}\frac{N^2}{m^2}k_1^2\mc{A}_1 + \frac{1}{4}\frac{N^2}{m^2}k_2^2\mc{A}_2,\\
		\mc{A}_f &=  \mc{A}_1 +  \mc{A}_2,
		\end{align}
	\end{subequations}
	where $A_i=|\hat{A}(k_i)|^2$ with $\hat{\cdot}$ the Fourier transform.
	
	Following the idea of \citet{Kraichnan1967} and \citet{Eyink1996}, by considering that 
	\begin{equation}\label{EA}
	\mc{E}_f = \frac{1}{4}\frac{N^2}{m^2}k_f^2\mc{A}_f,
	\end{equation}
	we can solve (\ref{balance1}) to obtain
	\begin{equation}
	\mc{A}_1 = \frac{k_f^2-k_2^2}{k_1^2-k_2^2}\mc{A}_f \quad\mathrm{and}\quad \mc{A}_2 = \frac{k_f^2-k_1^2}{k_2^2-k_1^2}\mc{A}_f.
	\end{equation}
	Then taking the limit $k_1\to0$ and $k_2\to\infty$ with fixed finite $k_f$, we obtain
	\begin{equation}\label{trans_dir0}
	\mc{A}_1 \to \mc{A}_f, \quad
	\mc{A}_2 \to 0,\quad
	\mc{E}_1 \to 0 \quad \mathrm{and} \quad
	\mc{E}_2 \to \mc{E}_f,
	\end{equation}
	which implies that the energy transfers downscale while the wave action transfers upscale. 
	
	The directions of energy transfer are opposite in the NIW-dominant and mean-flow-dominant cases.
	This implies that when the strength of NIW is intermediate energy can transfer both upscale and downscale simultaneously, which is consistent with the energy transfer scenario in the oceanic flows. In \S \ref{sec_num} we numerically check the existence of and study the details of the bidirectional energy transfer.
	
	To perform numerical simulations, the original YBJ-QG coupled model (\ref{NIWQG0}) is not exercisable due to a ``ultraviolet catastrophe": the dispersion term in (\ref{YBJ0}) implies that the resolved time scale behaves as the inverse of the square of wavenumber, which tends to infinity as the resolution increases.
	Therefore we consider running numerical simulations of the YBJ$^+$-QG coupled model proposed by \citet{Asselin2019}, which improves the numerical efficiency by retaining specific high-order terms and using the reconstitution technique to bound the frequency as wavenumber increases. 
	\citet{Asselin2019} derive the modified YBJ$^+$ equation but the form of wave-mean flow coupling is proposed. 
	We present a variational derivation that mimics the procedure of deriving the YBJ-QG model in \S \ref{sec_VP_YBJ+}. 
	And to ensure the validity of using the YBJ$^+$-QG model to study the transfer of conserved quantities, we check the argument of transfer direction in YBJ$^+$-QG coupled model in \S \ref{sec_trans_YBJ+}.

	\section{Numerical simulation}\label{sec_num}

To explore the transfer of conserved quantities in steady turbulent states, we add external forcing and artificial at both large and small scales to the inviscid YBJ$^+$-QG model
\begin{subequations}\label{NIWQG+}
	\begin{align}
	M_t + \Ja{\psi}{M}   + \frac{\ii f}{2}\frac{\nabla^2}{-\dfrac{f^2}{N^2}m^2+\frac{1}{4}\nabla^2}M + \frac{\ii}{2}M\nabla^2\psi  &= 0,\label{YBJ+}\\
	q_t + \Ja{\psi}{q} & = 0,\\
	\mathrm{with}\quad q = \nabla^2 \psi + \frac{\ii f}{2}\Ja{M^*}{M} + \frac{f}{4}\nabla^2 |M|^2,
	\end{align}
\end{subequations}
whose three-dimensional version is proposed by \citet{Asselin2019}. And we variaionally derive this coupled model in \S \ref{sec_VP_YBJ+}.  

Since (\ref{NIWQG+}) is derived from a variational approach, adding dissipation is artificial. 
We assume that the inviscid mechanism is not changed by the dissipations and the dissipations monotonically damp the total energy and wave action, so we choose to add dissipations in the equations of $M$ and mean vorticity $\nabla^2\psi$ instead of $q$. We will discuss the choice of dissipation terms in details below.
We obtain our model for numerical simulation:
\begin{subequations}\label{NIWQG_num}
	\begin{align}
	\pa{t}M + \Ja{\psi}{M} + \mc{P}M + \frac{\ii}{2}\nabla^2\psi M&= \alpha \nabla^{-2} M + \nu \nabla^6 M + R \frac{m}{Nk_f} F_1, \label{M_eq}\\
	\pa{t}\nabla^2\psi + \Ja{\psi}{\nabla^2\psi} + N(\psi,M) &= \alpha \psi + \nu \nabla^8 \psi + F_2, \label{Q_eq}
	\end{align}
\end{subequations}
where
\begin{equation}\label{N_express}
\begin{aligned}
N(\psi,M) = &\frac{f^2}{4}\sbr{\Ja{M^*}{\mc{P}M}-\Ja{\mc{P}M^*}{M}} + \frac{\ii f^2}{8}\nabla^2\br{ M\mc{P}M^* - M^*\mc{P}M}\\
&- \frac{f}{2}\nabla\cdot\Ja{\nabla\psi}{|M|^2}
\end{aligned}
\end{equation}
with 
\begin{equation}
\mc{P} = \frac{\ii f}{2}\frac{\nabla^2}{-\dfrac{f^2}{N^2}m^2+\frac{1}{4}\nabla^2}.
\end{equation}
Here, in the forcing term, ${m}/(Nk_f)$ is a normalized coefficient, $F_i$ (i=1 or 2) are temporal white-noise external forcing that is centred around wavenumber $|\bk|=k_f$, and $R$ is a tuning parameter that controls the relative strength of energy injection into the NIW and mean-flow components.
And when $R=1$, the ratio between the wave energy injection and QG-mean flow energy injection is $1/2$. 
Here, we add both hyper- and hypo-viscosities to dissipate the conserved quantities transferred downscale and upscale, respectively.

In the governing equations (\ref{NIWQG_num}) we design the forcing and dissipation in the mean vorticity equation instead of the PV equation to achieve that (i) if there is no external forcing and dissipation the conserved quantities are preserved, and (ii) the effect of viscous terms damp total energy and wave action, i.e. there is no viscous generation of the total energy.
However, as to the potential enstrophy, viscous generation is nonzero.
In other words, we choose to damp two out of three preserved quantities and focus on the fluxes of these two quantities.
It is debatable that this choice is realistic since we did not derive the dissipation terms from the original hydrostatic Boussinesq equations.
A potential way to achieve a realistic-scenario study is deriving the dissipative coupled model from viscous primitive equations, e.g. following the derivation in \cite{Rocha2018t}. 
However, the physical meaning of large-scale hypoviscosity is uncertain.
So considering the balance between the involute derivation and the aim of studying key mechanisms, we choose the above simple design as a heuristic starting point to understand the wave-mean flow coupled turbulence.

The numerical simulations use a Fourier pseudospectral method with 2/3 dealiasing in space, a resolution $512\times512$ in a domain of size $2\pi\times2\pi$ and a fourth-order explicit Runge--Kutta scheme in time, in which the nonlinear terms are treated explicitly, and linear terms implicitly use an integrating factor method. 
We take the forcing wavenumber to be $k_f=32$ and control the energy injection rate by the external forcing $F_i$ as $10^{-3}$, by choosing $\alpha=0.01$ and $\nu=10^{-12}$ we obtain small- and large-dissipation wavenumbers as $k_\alpha\approx0.4$ and $k_\nu\approx 100$, respectively.
We choose $f=1$, $N=1$ and $m=32$, therefore the forcing Burger number is $\Bu_f=1/4$ which is much smaller than $4$, and we expect the turbulent dynamics of conserved quantities transfer for the YBJ$^+$-QG system recover that for the YBJ-QG system (cf. \S \ref{sec_trans_YBJ+}).

\subsection{Dependence on the parameter $R$}

In this section, we focus on the dependence of the fluxes of energy and wave action in the spectral space on the parameter $R$.
Here, the spectral energy flux $F_E$ and wave action flux $F_A$ are defined from the equations
\begin{subequations}
	\begin{align}
	\pa{t}E(K) &= -\pa{K} F_E + \mathrm{forcing\, and\, dissipation},\\
	\pa{t}A(K) &= -\pa{K} F_A + \mathrm{forcing\, and\, dissipation},
	\end{align}
\end{subequations}
where conferring to (\ref{conserved_quantities}) we define
\begin{equation}
\int_{0}^{\infty} E(K) \dd K = \mc{E} \quad \mathrm{and} \quad  
\int_{0}^{\infty} A(K) \dd K = \mc{A}.
\end{equation} 
Here, $K=\abs{\bk}$ is the magnitude of the wavenunmber. 
And $F_E$ and $F_A$ indicate the energy and wave action transfer from small to large wavenumbers in the spectral space, respectively.

\begin{figure}
	\centering
	\includegraphics[width=0.49\linewidth]{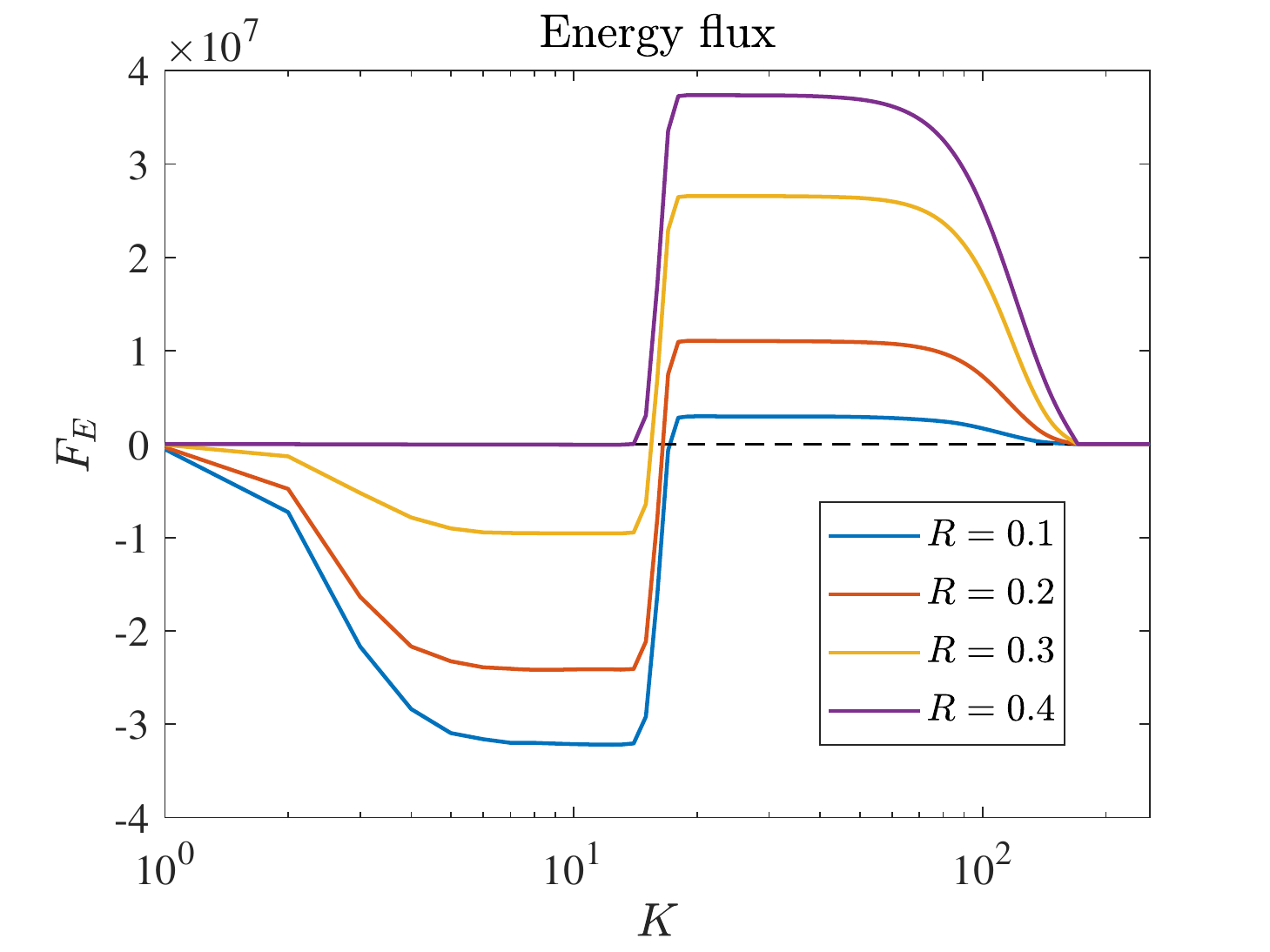}
	\includegraphics[width=0.49\linewidth]{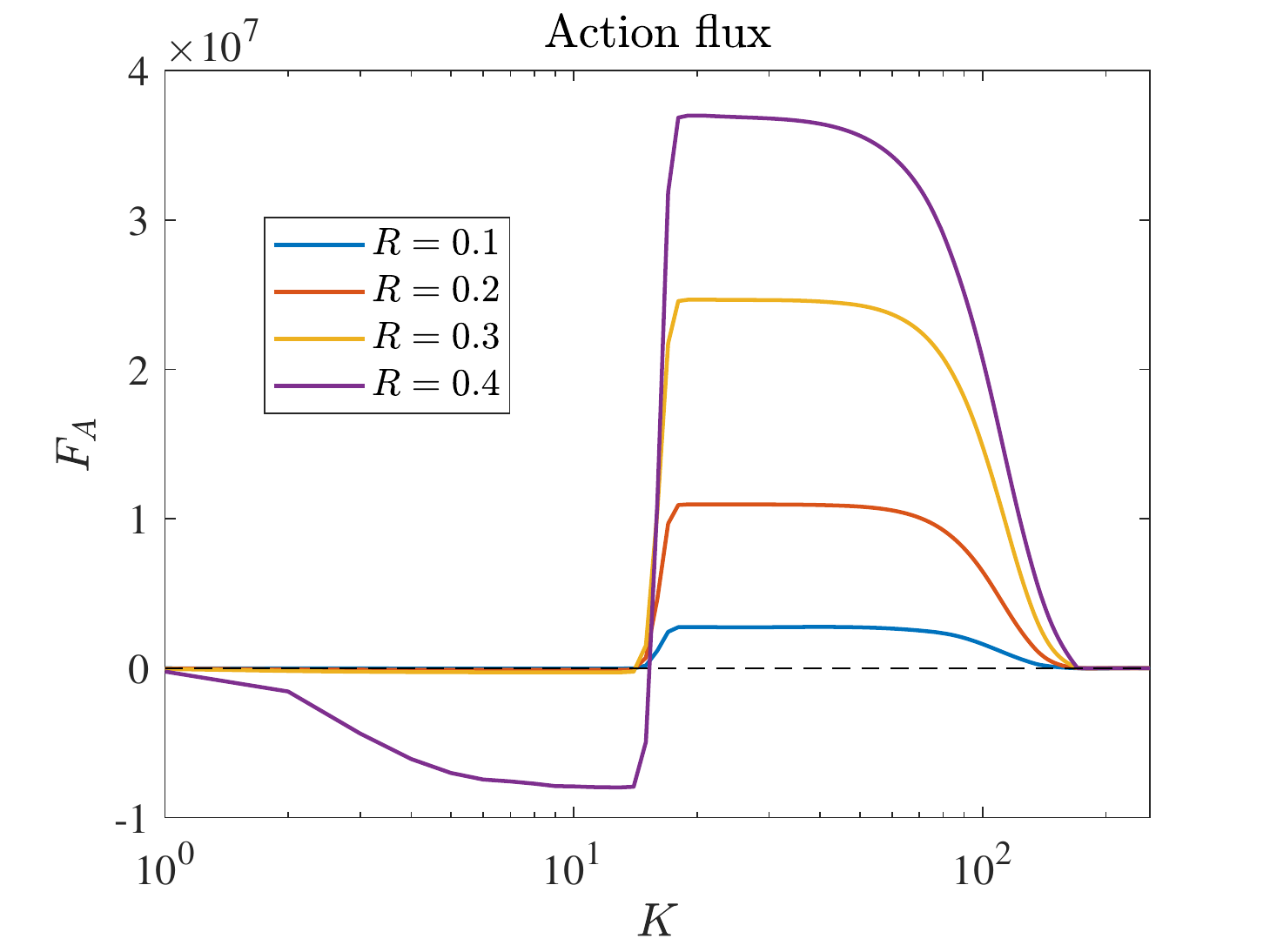}
	\caption{Dependence of the total energy flux and the wave action flux on parameter $R$, with $R=0.1,\,0.2,\,0.3$ and $0.4$. The negative and positive values of $F_E$ represent upscale and downscale energy transfers, respectively. The black dashed line is $0$ for reference. }
	\label{fig_e_flux_lam1024}
\end{figure}

In figure \ref{fig_e_flux_lam1024}, we show the dependence of total energy flux $F_E$ and wave action flux $F_A$ on parameter $R$. 
We observe the transition of energy transfer from upscale to bidirectional to downscale, which justifies our conjecture based on the argument in \S \ref{sec_arguement}.
In the corresponding parameter regimes, the wave action first transfers downscale then bidirectionally. 

\begin{figure}
	\centering
	\includegraphics[width=0.49\linewidth]{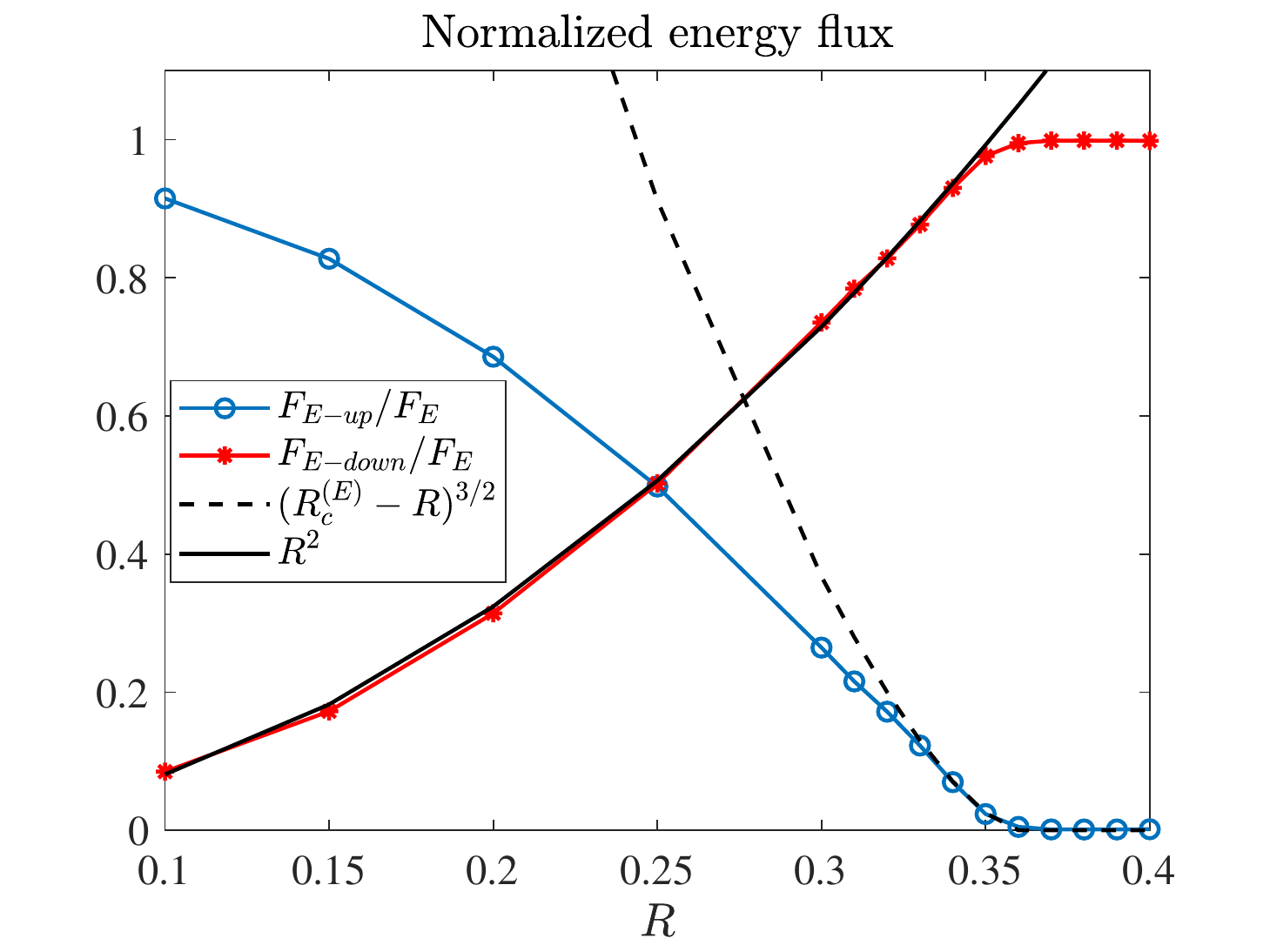}
	\includegraphics[width=0.49\linewidth]{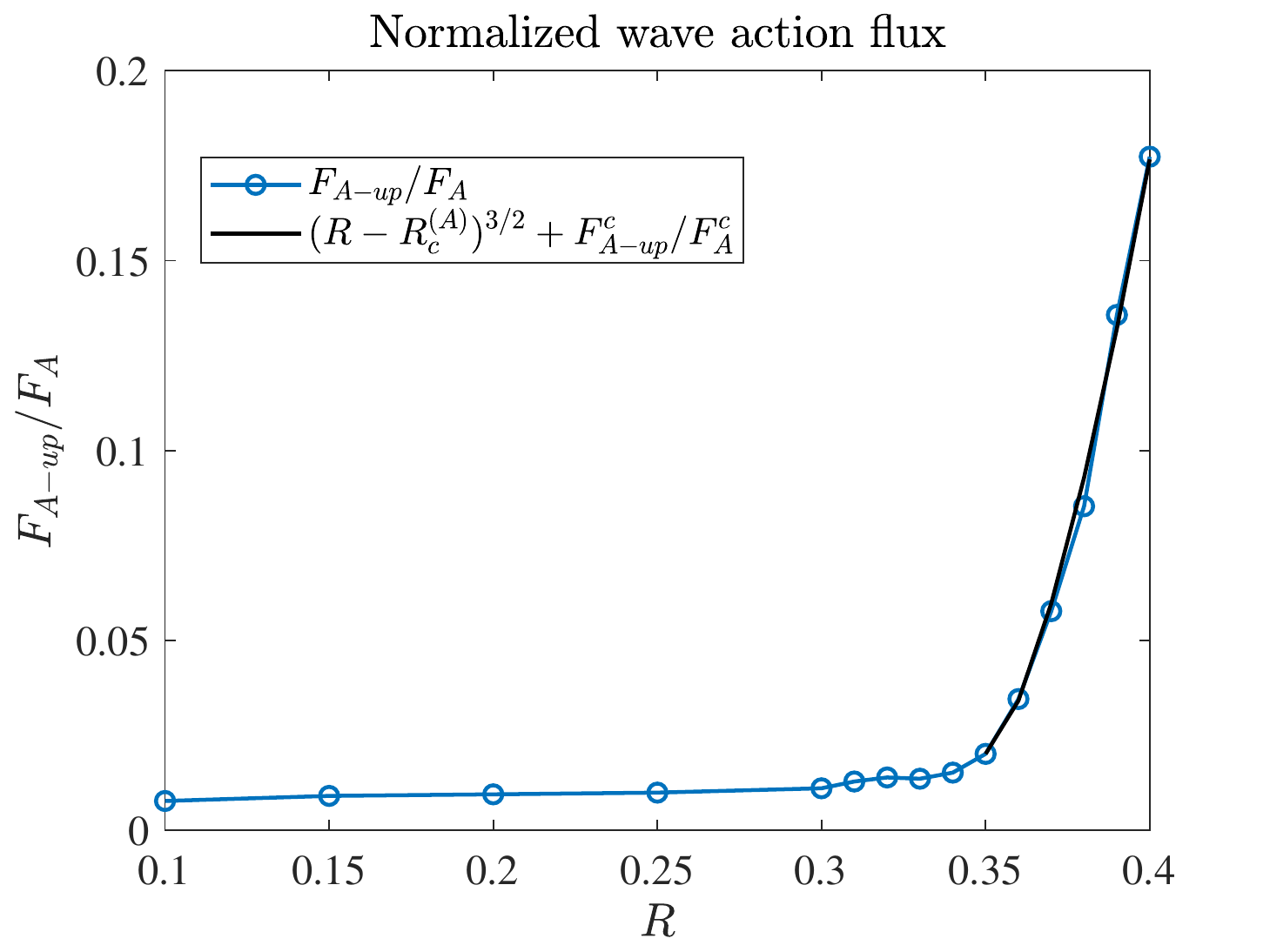}
	\caption{The dependence of normalized energy and wave action flux on the parameter $R$. Here, $R^{(E)}_c=0.36$ and $R^{(A)}_c=0.35$. The indices $E$ and $A$ represent the energy and action, respectively. And the lower indices $up$ and $down$ denotes the upscale and downscale fluxes, respectively. The index $c$ represents the critical point.}
	\label{fig_flux_R_nor}
\end{figure}

To show the details of the $R$-dependence, in the left panel of Figure \ref{fig_flux_R_nor}, we show the $R$-dependence of normalized upscale and downscale energy fluxes, which are defined as the upscale and downscale energy fluxes divided by the total energy injection, respectively.
We find a critical value of $R^{(E)}_c\approx0.36$. When $R<R_c$, the normalized upscale energy transfer monotonously decrease from $1$, which corresponds to the total upscale energy transfer in 2D turbulence, and when $R>R_c$ the normalized upscale energy transfer equals to zero.   
When $R<R^{(E)}_c$ the dependence of normalized downscale energy transfer scales as $R^2$, and close the critical value we find that the normalized upscale energy transfer scales as $(R^{(E)}_c-R)^{3/2}$, implying a second-order phase transition.

As to the normalized upscale action transfer shown in the right panel of Figure \ref{fig_flux_R_nor}, we find a critical value $R^{(A)}_c\approx0.35$. 
When $R<R^{(A)}_c$, the normalized upscale action transfer is almost zero, which may be interpreted as a passive-scalar-like wave magnitude; when $R>R^{(A)}_c$, the normalized upscale action transfer monotonously increases.
Around the critical value, the normalized upscale action transfer scales as $(R-R^{(A)}_c)^{3/2}$, indicating a second-order phase transition.
Here, due to the finite dissipation, when $R<R^{(A)}_c$, the normalized upscale action transfer is small but nonzero, so when fitting the power function near the critical point, we subtracted the small nonzero value at the critical point.
Based on the normalized transfers we find two close critical values $R^{(E)}_c\approx0.36$ and $R^{(A)}_c\approx0.35$, but we do not know if they are the same so we keep two symbols for them.

\subsection{Simulation with $R=0.3$} \label{sec_R03}

To compare the difference between the turbulent states across the critical point, we study in detail two cases with $R<R_c$ and $R>R_c$ in this and next subsections, respectively.
Here, since $R^{(E)}_c$ and $R^{(A)}_c$ are close, so, e.g., we use the notation $R_c$, and by $R<R_c$ we mean $R$ is smaller than both $R^{(E)}_c$ and $R^{(A)}_c$.
In this subsection, we show the details of the simulation with $R=0.3<R_c$.

\begin{figure}
	\centering
	\includegraphics[width=0.49\linewidth]{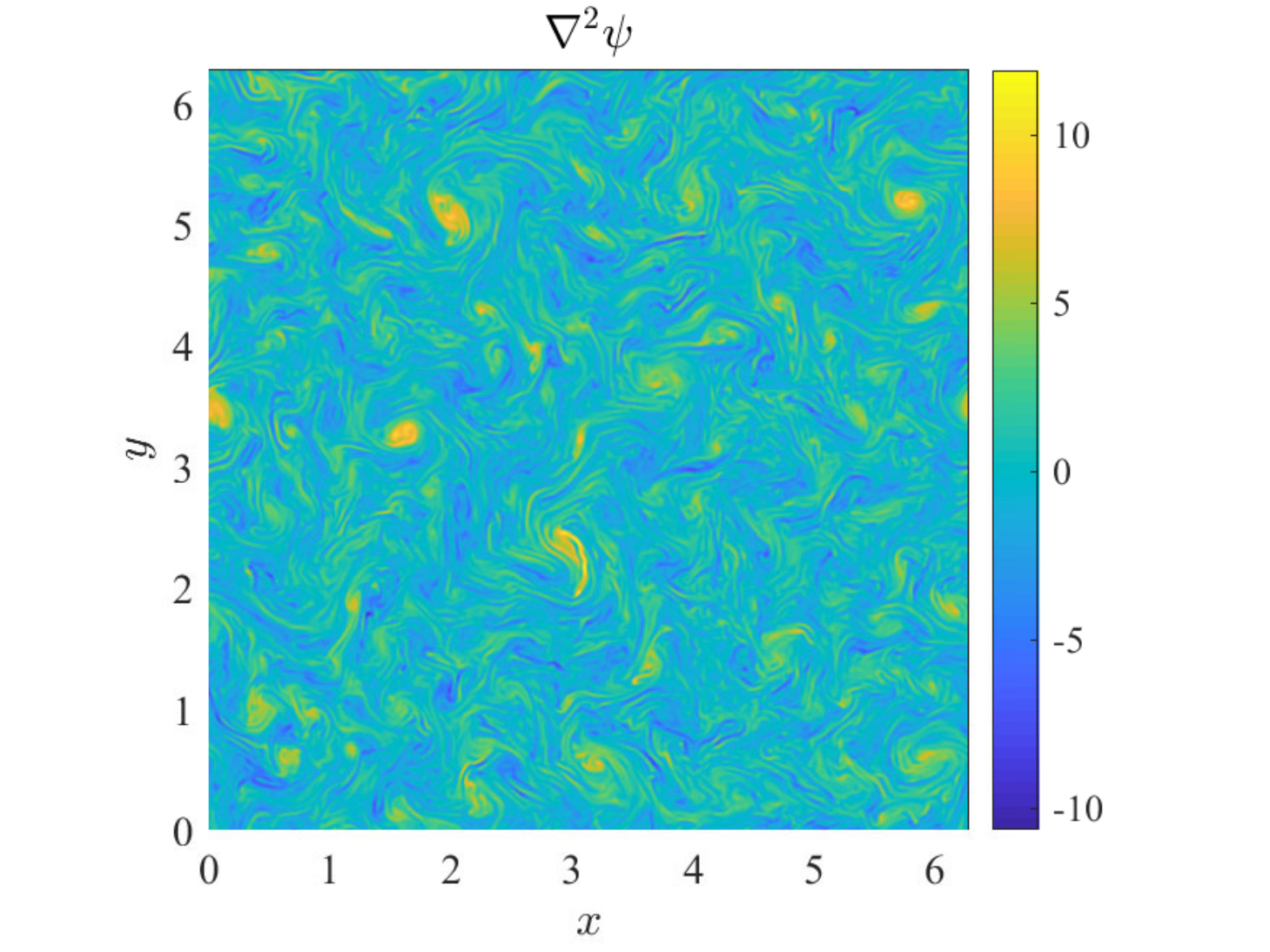}
	\includegraphics[width=0.49\linewidth]{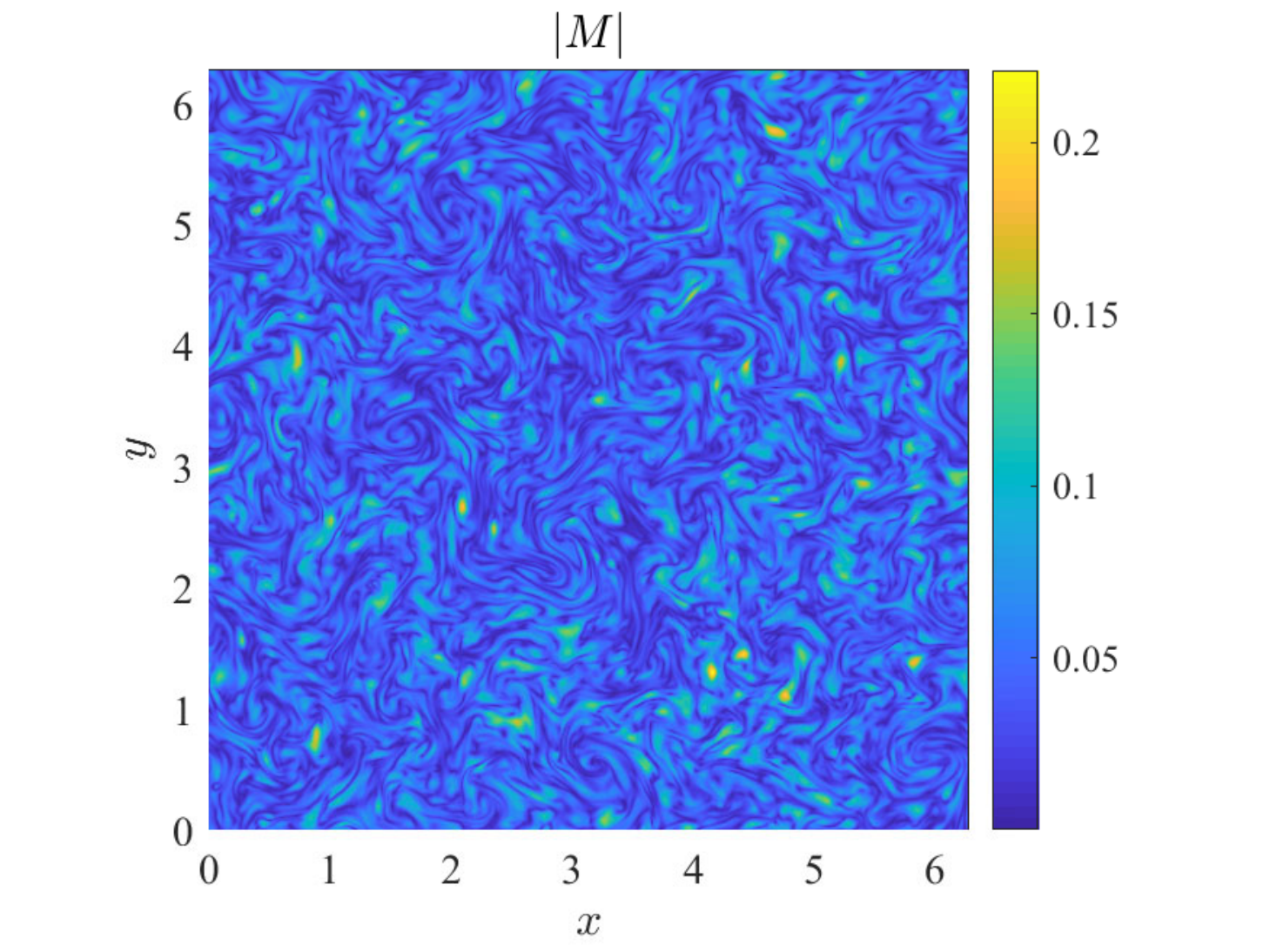}
	\caption{Snapshots of mean-flow vorticity and the wave amplitude at the turbulent statistically steady state in the simulation with $R=0.3$. }
	\label{fig_snap_R03}
\end{figure}

We show the snapshots of the of mean-flow vorticity and the wave amplitude at a turbulent statistically steady state in Figure \ref{fig_snap_R03}. 
There are more strong cyclones observed compared with the anticyclone, and the waves are advected by the mean flow.  
NIW's concentration at the anticyclone could explain the observed strong cyclones \citep{Danioux2015} since the anticyclone tend to be destroyed by the interaction with NIWs.

\begin{figure}
	\centering
	\includegraphics[width=0.49\linewidth]{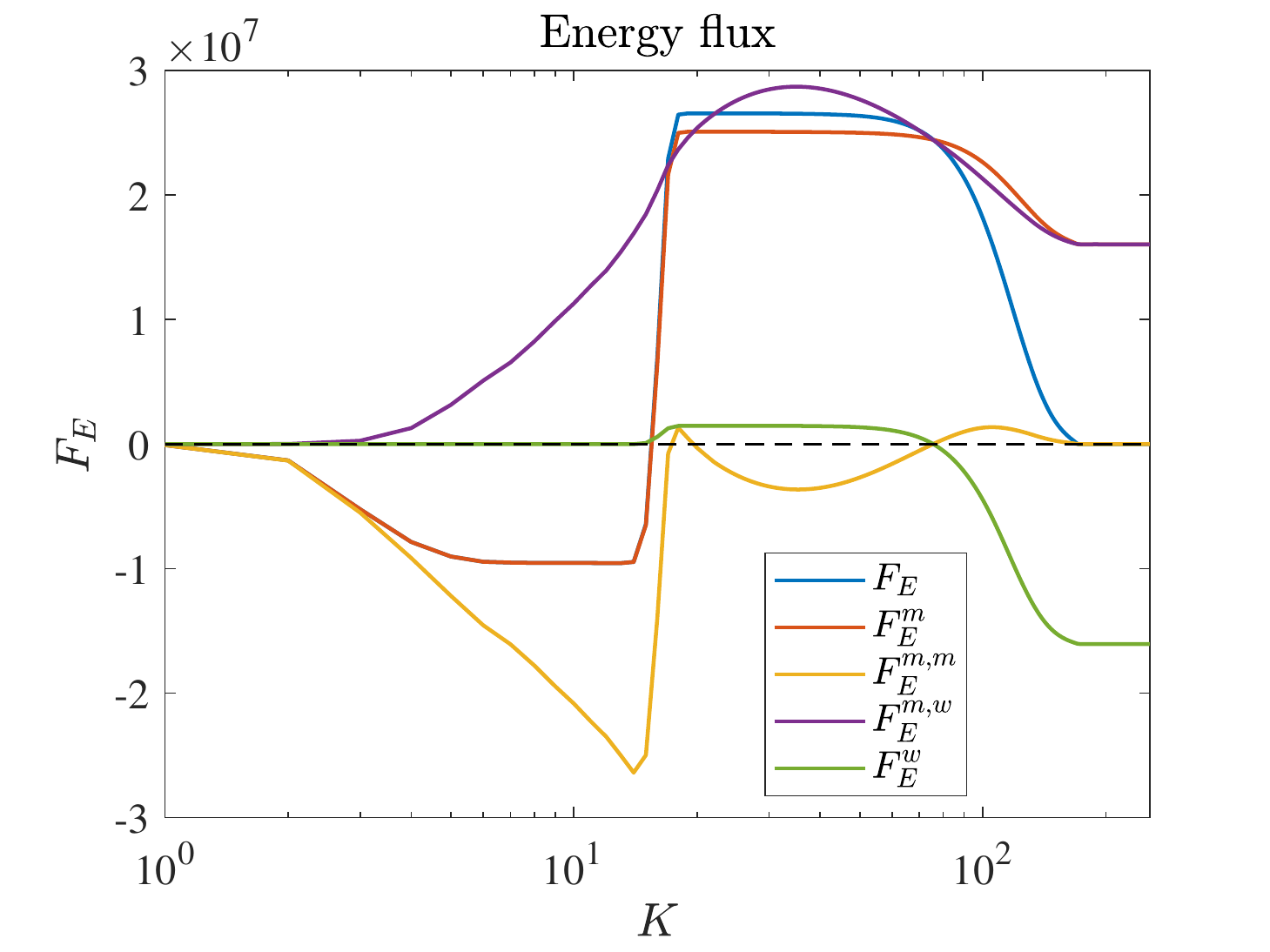}
	\includegraphics[width=0.49\linewidth]{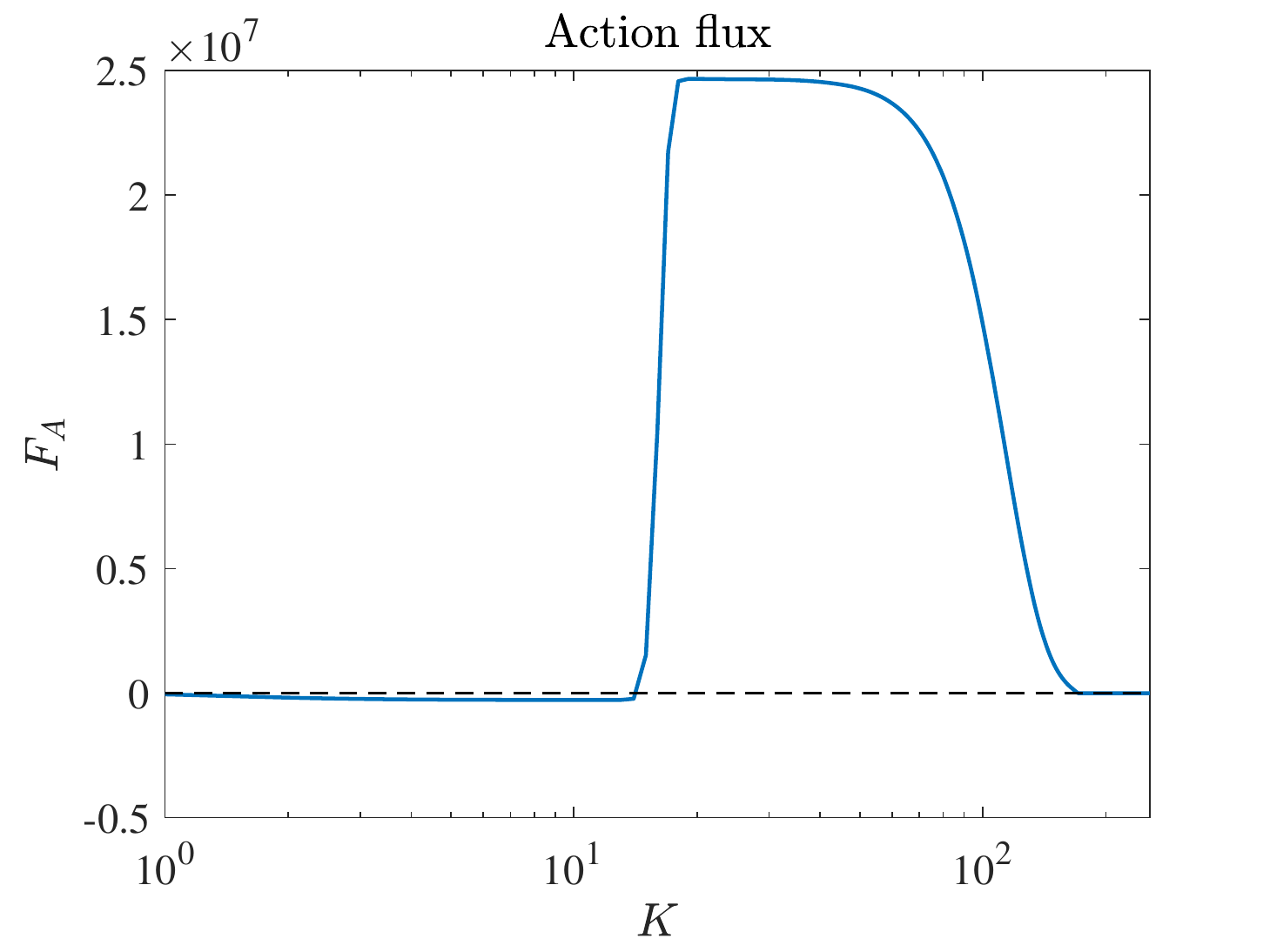}
	\caption{Energy and wave action flux for simulation with $R=0.3$. $F_E$ is the total energy flux; $F_E^w$ is the wave energy flux; $F_E^{m,m}$ and $F_E^{m,w}$ are the mean-flow-induced and NIW-induced mean energy fluxes, respectively. 
		All the quantities are normalized by the total energy injection $\epsilon$.
		}
	\label{fig_flux_R03}
\end{figure}

We show the different components of energy flux in the left panel of Figure \ref{fig_flux_R03}. 
We decompose the total energy transfer $F_E$ into the mean-flow component $F_E^m$ and the wave component $F_E^w$, which are calculated from the mean equation (\ref{Q_eq}) and the wave equation (\ref{M_eq}), and they correspond to the transfer of QG mean-flow energy and NIW energy (\ref{energy_decomp}), respectively.
Also, in the transfer of mean energy we distinguish the contributions from mean effect, $\Ja{\psi}{\nabla^2\psi}$, and the wave-mean flow interaction, $N(\psi,M)$, in the mean equation (\ref{M_eq}), and name them $F_E^{m,m}$ and $F_E^{m,w}$, respectively. 
It is observed that the wave energy transfers downscale with constant flux. The mean energy transfers bidirectionally where the mean flow dominantly induces the upscale energy transfer while the NIW dominantly induces the downscale energy transfer. 
Above the forcing scale, these two effects complete and result in a residue of upscale energy transfer. 
The wave action flux is shown in the right panel of Figure \ref{fig_flux_R03}. As we discussed above that the NIW almost behaves as a passive scalar, the wave action transfers downscale. 

\subsection{Simulation with $R=0.4$} \label{sec_R0301}

In this section, we show details of the numerical simulation with $R=0.4>R_c$.

\begin{figure}
	\centering
	\includegraphics[width=0.49\linewidth]{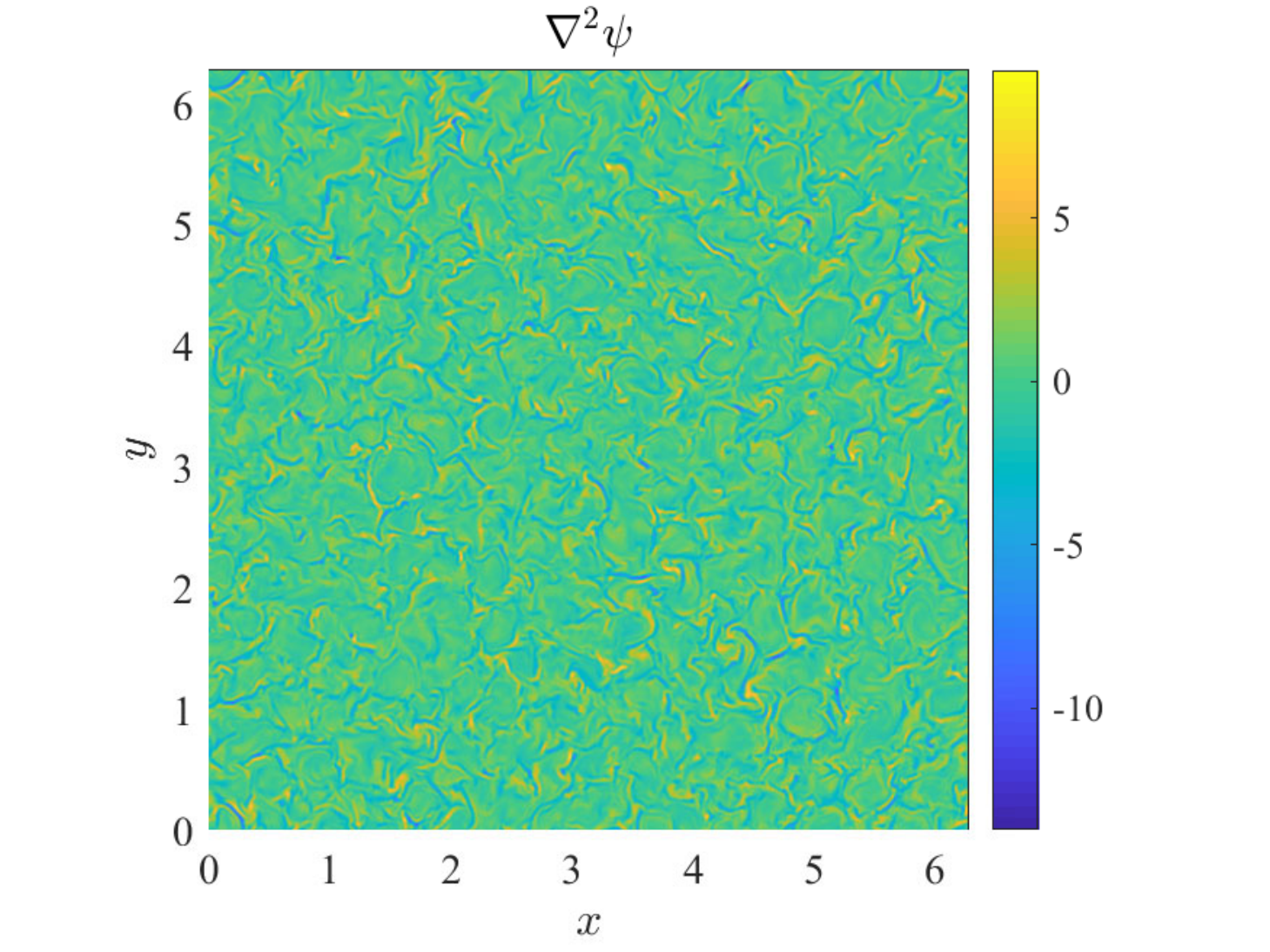}
	\includegraphics[width=0.49\linewidth]{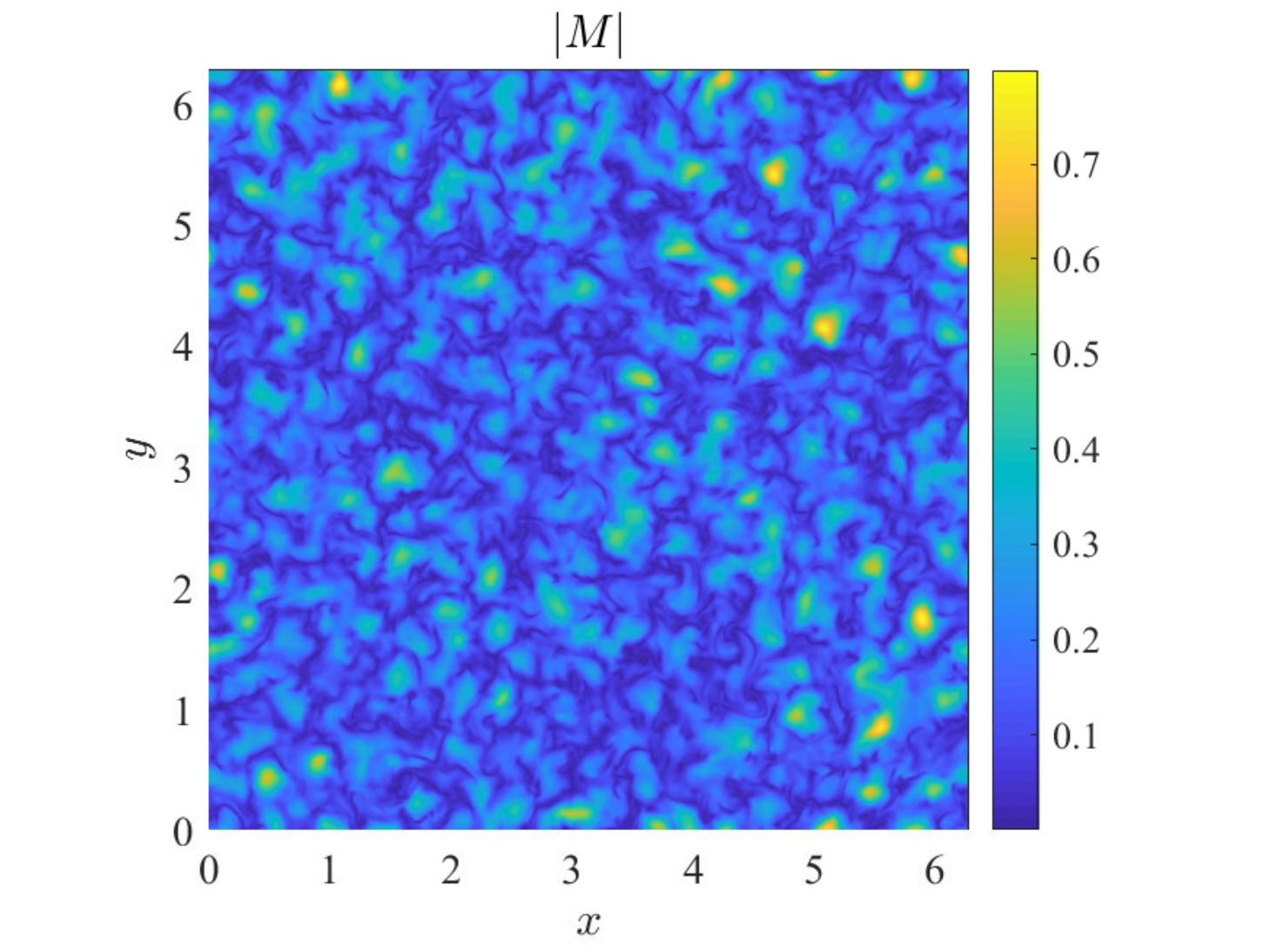}
	\caption{Snapshots of mean-flow vorticity and the wave amplitude at $t=2000$ in the simulation with $R=0.4$. }
	\label{fig_snap_R04}
\end{figure}

The snapshots of mean-flow vorticity and the wave amplitude at a turbulent statistically steady states are shown in Figure \ref{fig_snap_R04}. 
Comparing with Figure \ref{fig_snap_R03}, in the present simulation the strong cyclone patches are broken and structures of the forcing scale are observed in both fields.

\begin{figure}
	\centering
	\includegraphics[width=0.49\linewidth]{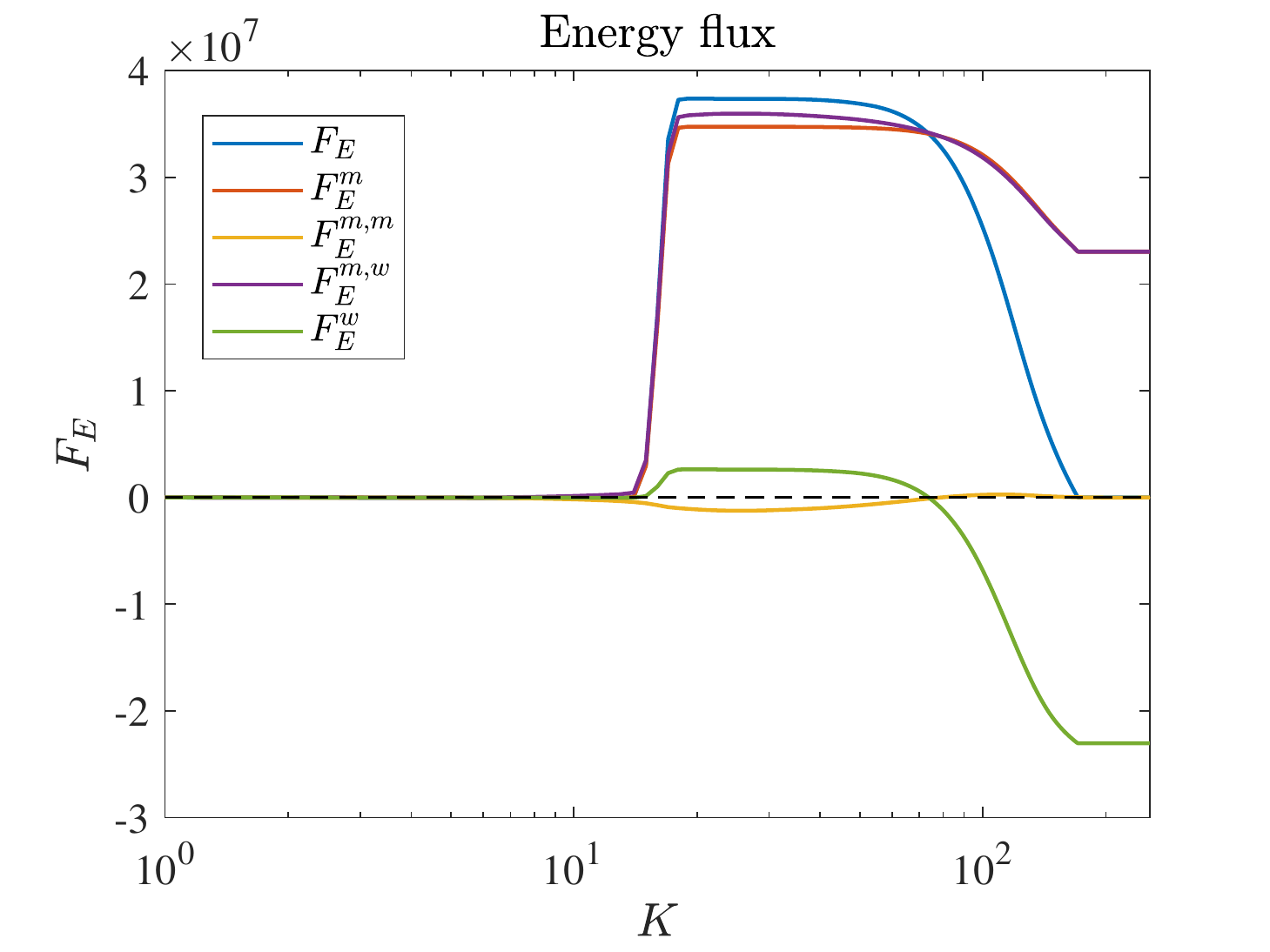}
	\includegraphics[width=0.49\linewidth]{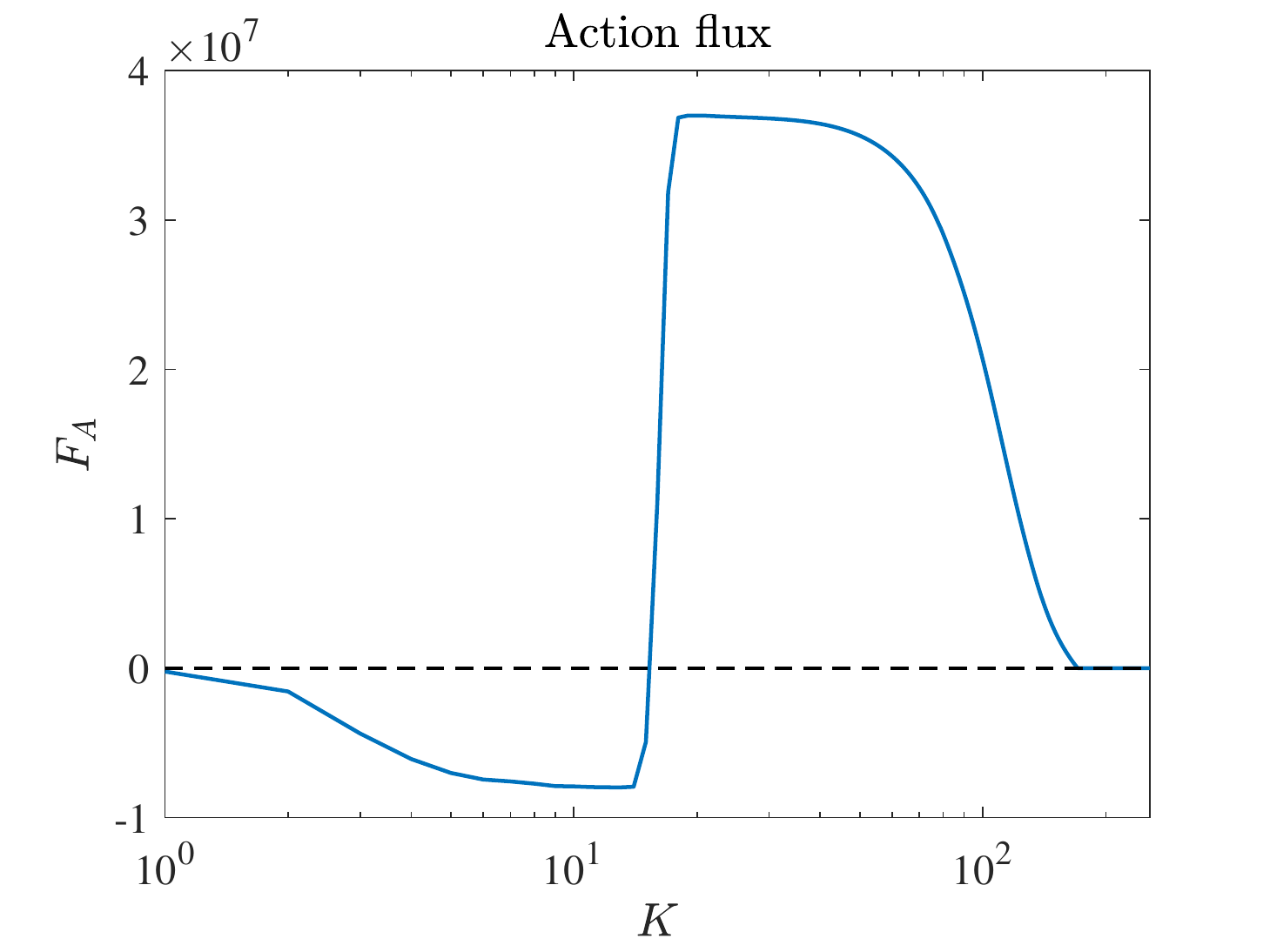}
	\caption{Energy and wave action flux for simulation with $R=0.4$. All notation are the same as those in Figure \ref{fig_flux_R03}.}
	\label{fig_flux_R04}
\end{figure}

The energy flux is shown in Figure \ref{fig_flux_R04}. We observe that the upscale fluxes are almost invisible, even though this is a 2D system with vorticity advection term, and the compensation between $F_E^{m,m}$ and $F_E^{m,w}$ above the forcing scale is not obvious. Now the wave action shows a clear bidirectional transfer.

\section{Discussion}\label{sec_dis}

\subsection{Mechanism of downscale mean energy transfer}\label{sec_CWI}

Using the model describing the interaction between NIW and QG mean flow \citep{Xie2015,Wagner2016}, the stimulated loss of balance (SLOB) (or stimulated NIW generation) mechanism, which explains the energy conversion from the QG mean flow to the NIW is proposed. 
Considering that the NIW transfers energy downscale and finally dissipate at the ocean interior, which is in the opposite energy transfer direction of mesoscale eddy itself, the SLOB is believed to be important for the energy balance puzzle of oceanic mesoscale eddies, even though \citet{AsselinYoung} recently find that the SLOB maybe not effective in the ocean using three-dimensional numerical simulations. 
However, the works mentioned above focus on initial-value problems, it remains to check whether the interaction between NIW and mesoscale eddies effectively impact the energetics of the latter in a statistically steady turbulent state, which is the topic of this paper.

We compare the states in the parameter regime $R<R_c$ with the SLOB mechanism, as in Figure \ref{fig_flux_R_nor} we find that in this regime the downscale energy flux is proportional to the injection of NIW energy, which is consistent with the SLOB (cf. \cite{Xie2015}).
However, the turbulent-state energy transfer mechanism differs from SLOB.
In the energy transfer of the $R=0.3$ simulation, Figure \ref{fig_flux_R03} shows that there is negligible energy conversion from the mean flow to the NIW, and the downscale energy transfer is dominated by the waves direct impact on the mean energy flux instead of the NIW's forward cascade.
In this turbulent downscale mean-energy transfer, the waves behave like a catalyst, so we name this downscale energy transfer mechanism as catalytic wave induction (CWI). 

Based on our toy-model simulation, CWI seems very effective. 
Around the critical value $R_c$, the ratio of the energy injection between wave and mean flow is $1/16$, i.e., the mean energy that changes the direction of energy transfer due to the NIW is $16$ times the injected wave energy, implying a much more substantial impact of the NIW on the direction of mean energy transfer compared with SLOB.
As we are focusing on turbulent statistically states, it may need to wait for a long time to establish this state. However, considering that wind has been blowing the ocean for a considerable long time, CWI is potentially crucial for oceanic mesoscale eddies.

Both SLOB and CWI show a $R^2$-dependence of downscale energy flux, where the former is explained based on the conserved quantities.
In CWI, this dependence can be understood from the nonlinear term (\ref{N_express}) where all terms depend on the quadratic of the wave magnitude.

We also need to note that CWI differs from the previous mechanism of downscale energy transfer in statistically steady states of NIW-mean flow interacting turbulence.
\citet{Barkan2017} study the direct extraction (DE) mechanism and SLOB in a wind-driven channel flow, where both mechanisms rely on energy conversion from mean flow to NIWs. 
But we cannot access the CWI's impact in their system as we need scale-by-scale energy transfer information to distinguish CWI from DE and SLOB.
In addition, the NIW's induction of downscale mean-flow energy flux is implicitly observed in \cite{Taylor2016}. 
They study the Reynolds stresses exerted by the near-inertial modes as the energy sink of the mean flow, and they access the strength of this energy sink in the spectral space.
After integrating their Fig. 5 over the wavenumber, we can find a NIW-induced downscale mean-energy flux.
But without separating and explicitly calculating the energy flux of mean flow and NIW, it is hard to access the importance of energy conversion between NIW and mean flow on the direction of mean flow energy transfer.


\subsection{Argument for the direction of energy transfer}

Our heuristic argument for transfer directions based on three conserved quantities (cf. \S \ref{sec_arguement}) is also applicable to other systems, such as the 2D magnetohydrodynamical (MHD) turbulence, where phase transition depending on the relative strength between the magnetic- and fluid-field forcing is observed \citep{Seshasayanan2014,Seshasayanan2016}.
In the 2D MHD system, the magnetic potential plays the role of the wave magnitude in the NIW-QG coupled system, and to apply our argument we need to replace the wave action by the square magnetic vector potential.
But in 2D MHD, the wave action is the Casimir (material invariant), in contrast to the PV in our wave-mean interaction system.
However, our argument can only qualitatively predict the existence of phase transition, and it cannot predict the type of phase transition and the existence of bidirectional energy transfer. 
The details of the phase transition depend on the specific forms of nonlinear terms, whose detailed mechanisms remain to be studied.  
We need to note that our NIW-QG system and the 2D MHD system are different in the details of energy transfer: in our NIW-QG system energy conversion from mean flow to the NIW is negligible, but in 2D MHD turbulence the energy conversion from kinetic energy to magnetic energy is important \citep{Seshasayanan2016}.

In both the NIW-QG system and the 2D MHD system, the transition of energy and wave action (square magnetic vector potential) are found to happen around the same critical value. Whether the two critical values are the same is not definitive, and we do not know yet the mechanism that identifies these critical values.  
Finding the detailed mechanism for this type of phase transition remains an open question.

It needs to note that the bidirectional energy transfer is not a result of the domain size, which differs from the bidirectional energy flux observed in energy flux-loop systems \cite[cf.][]{Boffetta2011,Falkovich2017,Xie2019}. 
E.g., in 2D inertia-gravity wave turbulence, the injected kinetic first cascades upscale then is converted to the potential energy which finally transfers downscale.
But for complete kinetic-potential energy conversion, the domain size should be larger than the Ozmidov scale, the scale energy conversion happens. 
If not, the incomplete energy conversion results in bidirectional energy transfer.
Here, our argument implies that even when the domain size is infinite, bidirectional energy transfer can still happen.

\subsection{Potential enstrophy flux}

In the above section, we pay very little attention to the potential enstrophy flux because the chosen artificial hyper- and hypo-viscosities in (\ref{NIWQG_num}) do not correspond to sign-definitive potential enstrophy ``dissipation".
This is because that the YBJ-QG model is initially derived in a variational framework, which is hard to include the dissipation effects, we have to design which quantities to dissipate. 
This situation is similar to that in shallow water model, where the two out of three conserved quantities are chosen to be dissipated \cite[cf.][]{Jacobson2008}.

\begin{figure}
	\centering
	\includegraphics[width=0.49\linewidth]{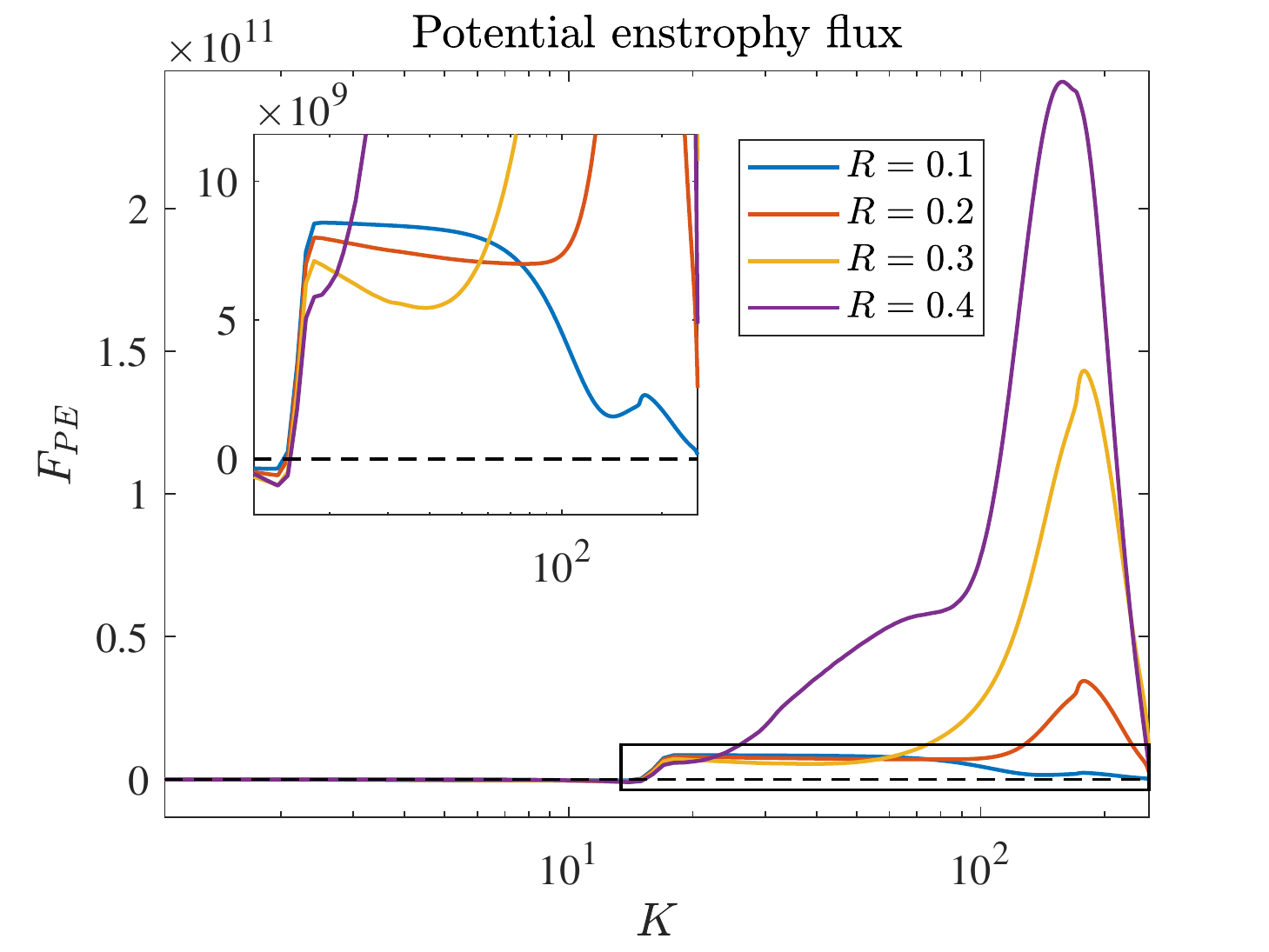}
	\caption{Potential enstrophy flux for simulation with $R=0.1,0.2,0.3$ and $0.4$. The inset zooms in the black-boxed region.}
	\label{fig_pe_flux}
\end{figure}

We present the $R$ dependence of the potential enstrophy flux in Figure \ref{fig_pe_flux}.
When $R$ is small, potential enstrophy transfers downscale since the weak wave implies the mean flow's enstrophy dominance of the potential enstrophy.
However, as $R$ increases viscous generation is not negligible and the potential energy fluxes are not constants. 

Since the YBJ-QG coupled model can also be derived in a Eulerian framework \cite[cf.][]{Wagner2016}, it is interesting to perform a derivation starting from the viscous Boussinesq equation to find the suitable dissipation terms \cite[cf.][Chapter 5, where the vertical viscosity is included]{Rocha2018t}. 

\section{Summary and conclusion} \label{sec_sum}


Using the NIW-QG coupled model (\ref{NIWQG_num}), we study the dependence of energy and wave action transfer directions on parameter $R$, which is defined as the ratio of the external forcing magnitude in the NIW and QG components. 
We propose a heuristic argument based on the inviscid preserved energy, potential enstrophy and wave action to predict the existence of phase transition, which is justified by numerical simulations. 
We find a critical value $R^{(E)}_c$ and $R^{(A)}_c$, across which the energy and wave action flux show a second-order phase transition, respectively.
Since $R^{(E)}_c$ and $R^{(A)}_c$ are close, they may be the same but we are not sure about it. But for simplicity, in the below summary we keep only one $R_c$.
(1) When $0<R<Rc$, the total energy transfers bidirectionally, and the normalized upscale energy flux monotonously decrease and reaches zero at $R=R_c$. The normalized upscale wave action flux remains a small close-to-zero value, so we believe the action transfers downscale.
The vorticity field consists of vorticity patches with cyclone dominance, but due to the wave impact more small scale structures are observed compared with the 2D turbulence. This is consistent with the information that even in 2D inertia-gravity waves can induce downscale energy transfer (cf. \cite{Xie2019}). These turbulent states also show the consistency with the concentration of NIW at the anticyclones \citep{Danioux2015} since the cyclones dominate the vorticity field.
(2) When $R>R_c$, energy transfers downscale; the wave action transfers bidirectionally, and as $R$ increases the upscale energy transfer increases vortex filaments are the dominant structure, which is similar to the 3D turbulence.

In the forced-dissipative turbulence of NIW-QG mean flow interaction, we discover a new mechanism -- catalytic wave induction (CWI), which is responsible for the downscale energy transfer. 
Different from the SLOB mechanism, where the QG mean energy converts to the wave energy and then cascades downscale, in CWI, the waves play a catalytic role, and the mean energy transfers downscale without recognizable conversion to the wave energy.

We close this paper by pointing out some potential future works relating to the energy transfer in the NIW-mean flow coupled system:
(i) Similar to the SLOB, the downscale energy flux in CWI is also proportional to the wave energy injection rate, and it is shown to be effective (cf. \ref{sec_CWI}) based on our toy model study, but we still cannot calculate the value of the constant linking the downscale energy flux and the wave energy injection rate.
We would like to further explore the dependence of the constant on parameters such as the Burger number.
(ii) This paper's toy-model study indicates that CWI is potentially important for the ocean energy energetics, and there are potential external-forcing-induced sudden changes in the direction of energy transfer in the ocean. These implications remain to be checked and justified in more realistic simulations and observations.
(iii) We need to note that the mechanism of NIW saturation differs from that in the wave turbulence theory \cite[cf.][]{Zakharov2012} because the governing equation (\ref{NIWQG0}) does not contain wave nonlinear term and the nonlinear wave interaction only through the mean flow generation. 
However, the conserved quantities in the wave dominant regime are the same as those in the wave turbulence theory, the transfer directions of conserved quantities are the same as that predicted by the (weak) wave turbulence theory. So it is our future work to study the difference and the link between these two mechanisms for turbulent wave saturation, which involves the present not-reached large-$R$ regime.\\

\noindent\textbf{Acknowledgment}

\noindent\textbf{Declaration of Interests}

The authors report no conflict of interest.

\appendix

\section{YBJ$^+$-QG coupled model} \label{sec_YBJ+}

\subsection{A variational derivation for the YBJ$^+$-QG model}\label{sec_VP_YBJ+}

In this section, we modify the variational framework in \cite{Xie2015} to obtain the YBJ$^+$-QG coupled model proposed by \citet{Asselin2019}.

The reconsitutation by Asselin \& Young modifies the dispersion relation of NIW as
\begin{equation}
\omega = f + \frac{4\Bu}{4+\Bu}\frac{f}{2}=f+\sigma^+, \label{dr}
\end{equation}
where the first $f$ is absorbed in the ansatz of NIW and the second part $\sigma^+$ should be obtained from the YBJ$^+$ equation describing the modulation of NIW amplitude.
Here we introduce the Burger number \[\Bu = \frac{N^2k^2}{f^2m^2}.\] 

The spirit of YBJ$^+$ equation motivates us to introduce the following mappings:
\begin{subequations}\label{map}
	\begin{align}
	\chi \to \mc{K} A,\\
	\pa{z} \to \frac{N}{f}\mc{K}, 
	\end{align}
\end{subequations}
where $u+\ii v = -\ii f\chi \ex^{-\ii ft}$ and
\begin{equation}
\mc{K} = \br{\frac{1}{2}\pa{x},\frac{1}{2}\pa{y},\frac{f}{N}\pa{z}},
\end{equation}
therefore $\mc{K}\cdot \mc{K}=L^+$. Here we have assumed that $f$ and $N$ are constants for simplicity.

Applying this mapping to the NIW Lagrangian (cf. (3.23) in \cite{Xie2015} and taking $\beta=0$) we obtain
\begin{equation}
\begin{aligned}
\av{\mc{L}}_{NIW} = -\int \left( \frac{\ii f}{4} \br{L^+A D_T L^+A^* - L^+A^* D_T L^+A} \right. \\ 
\left. + f\psi G(\mc{K}A^*,\mc{K}A) + \frac{1}{4}N^2 |\nabla \mc{K}A|^2  \right) \dd \bx ,
\end{aligned}
\end{equation} 
where $D_T=\pa{T} + \Ja{\psi}{\cdot}$ and 
\begin{equation}
G(\mc{K}A^*,\mc{K}A) = \frac{1}{2}\br{ 2|\nabla \mc{K} A_z|^2 - \mc{K}A_{zz}\nabla^2\mc{K}A^* - \mc{K}A^*_{zz}\nabla^2\mc{K}A}.
\end{equation}

Taking variation of $A^*$ we obtain the wave equation
\begin{equation}\label{YBJ+1}
L^+\br{D_T(L^+A)} + \frac{\ii f}{2} L^+\nabla^2 A + \frac{2\ii}{f}L^+\br{ \nabla\cdot\br{\psi\nabla L^+A} - \frac{1}{2}\mc{K}\cdot\br{\psi\nabla^2\mc{K} A} } + \frac{1}{2} \nabla^2\mc{K}\cdot\br{\psi \mc{K}^{3/2}A} = 0.
\end{equation}	
Note that (\ref{YBJ+1}) has three refraction terms that are comparable with that in the first YBJ equation \citep{Young1997}.

We can multiply $(L^+)^{-1}$ to (\ref{YBJ+1}) to obtain a complicated version of the YBJ$^+$ equation:
\begin{equation}\label{YBJ+2}
D_T(L^+A) + \frac{\ii f}{2} \nabla^2 A + \frac{2\ii}{f} \nabla\cdot\br{\psi\nabla L^+A} - \frac{\ii}{f}\mc{K}\cdot\br{\psi\nabla^2\mc{K} A}  + \frac{1}{2} \nabla^2\mc{K}^{-1}\cdot\br{\psi \mc{K}^{3/2}A} = 0,
\end{equation}
where operator $\mc{K}^{-1}$ is nonlocal in the last refraction term. 

But for the modeling purpose, we introduce a modified wave potential energy
\begin{equation}
G\to G^+ = \frac{1}{4}\nabla^2|L^+A|^2,
\end{equation}
corresponding to the vertical averaging procedure introduced by \citet{Wagner2016}, 
then the Lagrangian of NIW become  
\begin{equation}
\begin{aligned}
\av{\mc{L}}_{NIW+} = -\int \left( \frac{\ii f}{4} \br{L^+A D_T L^+A^* - L^+A^* D_T L^+A} + \frac{f}{4}\psi\nabla^2|L^+A|^2 + \frac{1}{4}N^2 |\nabla \mc{K}A|^2  \right) \dd \bx,
\end{aligned}
\end{equation}
then taking the variation of $A^*$ we recovers the YBJ$^+$ equation:
\begin{equation}\label{YBJ+a}
D_T(L^+A) + \frac{\ii f}{2} \nabla^2 A + \frac{\ii}{2}\nabla^2\psi L^+A = 0.
\end{equation}

Thus, the potential vorticity (PV) with barotropic mean flow becomes
\begin{equation}
\begin{aligned}
q = &\nabla^2\psi +\frac{f^2}{N^2}\psi_{zz}  + \frac{\ii f}{2}\Ja{L^+A^*}{L^+A} + fG^+\\
= &\nabla^2\psi +\frac{f^2}{N^2}\psi_{zz}  + \frac{\ii f}{2}\Ja{L^+A^*}{L^+A} + \frac{f}{4}\nabla^2|L^+A|^2.
\end{aligned}
\end{equation}
The relabeling symmetry implies the PV conservation
\begin{equation}
q_t + \Ja{\psi}{q} = 0,
\end{equation} 
that together with the YBJ$^+$ equation (\ref{YBJ+a}) form a closed couple system.
Then, assuming that the mean flow is barotropic, i.e. $\pa{z}=0$, and the wave has only one vertical wave wavenumber $m$, and renaming the variable $L^+A$ as $M$, we obtain the YBJ$^+$-QG model (\ref{NIWQG+}).

\subsection{Argument for the transfer of conserved quantities in YBJ$^+$-QG model}\label{sec_trans_YBJ+}

Note that instead of the variable $A$ used by \citet{Asselin2019} in the above equations we introduce $M=\ii L^+A /f$, where $L^+A = -f^2m^2/N^2 + \nabla^2/4$, to directly compare (\ref{NIWQG+}) with (\ref{NIWQG0}).  
We can see that comparing with (\ref{YBJ0}) the YBJ$^+$ equation (\ref{YBJ+}) modifies the linear dispersion term, which bounds the maximum frequency to be resolved as $2f$ instead of $\infty$, thus, the numerical simulations are accelerated. 
Meanwhile, the nonlinear terms are not changed, making the system (\ref{NIWQG+}) remains a good model to study the nonlinear dynamics of NIW-mean flow interaction.

Same as the YBJ-QG model (\ref{NIWQG0}), the YBJ$^+$-QG model (\ref{NIWQG+}) preserves the total energy, potential enstrophy and the wave action 
\begin{subequations}
	\begin{align}
	\mc{E} &= \int \br{\frac{1}{2}|\nabla \psi|^2 + \frac{f^2}{4}|\mc{L}M|^2} \dd \bx,\label{energy}\\
	\mc{P} &= \int  q^2 \dd \bx,\\
	\mc{A} &= \int |M|^2\dd \bx,
	\end{align}
\end{subequations}
where \[\mc{L}^2 = \frac{\nabla^2}{\dfrac{f^2}{N^2}m^2-\frac{1}{4}\nabla^2}.\]

In the wave-dominant case, the energy and (potential) enstrophy dominates the dynamics, so the energy transfers upscale and the (potential) enstrophy transfers downscales. Since the waves are nearly passive, the wave action should transfer downscale.

In the mean-flow-dominant case, energy and wave action control the dynamics, and their leading order expressions are
\begin{subequations}
	\begin{align}
	\mc{E} &= \int \frac{f^2}{4}|\mc{L}M|^2 \dd \bx,\\
	\mc{A} &= \int |M|^2\dd \bx.
	\end{align}
\end{subequations}
Again we consider the heuristic scenario that the energy and wave action are injected at an intermediate wavenumber $k_f$ and dissipate at both small and large wavenumbers $k_1$ and $k_2$, therefore the conserved energy and wave action imply
\begin{subequations}\label{balance2}
	\begin{align}
	\mc{E}_f &= \frac{f^2}{4}\frac{k_1^2}{\dfrac{f^2}{N^2}m^2+\frac{1}{4}k_1^2}\mc{A}_1 + \frac{f^2}{4}\frac{k_2^2}{\dfrac{f^2}{N^2}m^2+\frac{1}{4}k_2^2}\mc{A}_2,\\
	\mc{A}_f &= \mc{A}_1 + \mc{A}_2.
	\end{align}
\end{subequations}
Considering the relation (\ref{EA}) we can solve (\ref{balance2}) to obtain
\begin{equation}
\mc{A}_1 = \frac{\br{4f^2m^2+N^2k_1^2}\br{k_f^2-k_2^2}}{\br{4f^2m^2+N^2k^2}\br{k_1^2-k_2^2}}\mc{A}_f \quad\mathrm{and}\quad 
\mc{A}_2 = \frac{\br{4f^2m^2+N^2k_2^2}\br{k_f^2-k_1^2}}{\br{4f^2m^2+N^2k^2}\br{k_2^2-k_1^2}}\mc{A}_f.
\end{equation}

In the limit $k_1\to0$ and $k_2\to\infty$, the leading order energy and wave action dissipation can be approximated as 
\begin{subequations}\label{express}
	\begin{align}
	\mc{A}_1 = \frac{4}{4+\Bu_f}\mc{A}_f,\quad
	\mc{A}_2 = \frac{\Bu_f}{4+\Bu_f}\mc{A}_f,\quad
	\mc{E}_1 = 0 \quad \mathrm{and} \quad
	\mc{E}_2 = \mc{E}_f,
	\end{align}
\end{subequations}
where $\Bu_f=N^2k_f^2/\br{f^2m^2}$ is the forcing Burger number. 
It is interesting to observe that the YBJ$^+$-QG model can transfers energy to both wave action to both large and small scales depending on the value of forcing Burger number.
If we want the energy and action transfer to recover those obtained in the YBJ-QG model (\ref{trans_dir0}), the forcing Burger number should be much smaller than $4$.

	\bibliographystyle{jfm}
	\bibliography{YBJref}

\begin{thebibliography}{37}
\expandafter\ifx\csname natexlab\endcsname\relax\def\natexlab#1{#1}\fi

\bibitem[Alexakis \& Biferale(2018)]{Alexakis2018}
{\sc Alexakis, A. \& Biferale, L.} 2018 Cascades and transitions in turbulent
  flows. {\em Phys. Rep\/} {\bf 767--769}, 1--101.

\bibitem[Andrews \& McIntyre(1978)]{Andrew1978}
{\sc Andrews, D.~G. \& McIntyre, M.~E.} 1978 Generalised {E}liassen-{P}alm and
  {C}harney-{D}razin theorems for waves on axisymmetric mean flows in
  compressible atmospheres. {\em J. Atmos. Sci.\/} {\bf 35}, 175--185.

\bibitem[Asselin \& Young(2019{\natexlab{{\em a\/}}})]{Asselin2019}
{\sc Asselin, O. \& Young, W.~R.} 2019{\natexlab{{\em a\/}}} An improved model
  of near-inertial wave dynamics. {\em J. Fluid Mech.\/} {\bf 876}, 428--448.

\bibitem[Asselin \& Young(2019{\natexlab{{\em b\/}}})]{AsselinYoung}
{\sc Asselin, O. \& Young, W.~R.} 2019{\natexlab{{\em b\/}}} Penetration of
  wind-generated near-inertial waves into a turbulent ocean. {\em arXiv\/} {\bf
  physics.ao-ph}, 1912.08323v1.

\bibitem[Barkan {\em et~al.\/}(2017)Barkan, Winters \& McWilliams]{Barkan2017}
{\sc Barkan, R., Winters, K.~B. \& McWilliams, J.~C.} 2017 Stimulated imbalance
  and the enhancement of eddy kinetic energy dissipation by internal waves.
  {\em J. Phys. Oceanogr.\/} {\bf 47}, 181--198.

\bibitem[Benavides \& Alexakis(2017)]{Benavides2017}
{\sc Benavides, S.~J. \& Alexakis, A.} 2017 Critical transitions in thin layer
  turbulence. {\em J. Fluid Mech.\/} {\bf 822}, 364--385.

\bibitem[Boffetta {\em et~al.\/}(2011)Boffetta, {De Lillo}, Mazzino \&
  Musacchio]{Boffetta2011}
{\sc Boffetta, G., {De Lillo}, F., Mazzino, A. \& Musacchio, S.} 2011 A flux
  loop mechanism in two-dimensional stratified turbulence. {\em Europhys.
  Lett.\/} {\bf 95}, 34001.

\bibitem[Cardy {\em et~al.\/}(2008)Cardy, Falkovich \& Gawedzki]{Cardy2008}
{\sc Cardy, J., Falkovich, G. \& Gawedzki, K.} 2008 {\em Non-equilibrium
  statistical mechanics and turbulence\/}. Cambridge University Press.

\bibitem[Danioux {\em et~al.\/}(2015)Danioux, Vanneste \&
  B\"uhler]{Danioux2015}
{\sc Danioux, E., Vanneste, J. \& B\"uhler, O.} 2015 On the concentration of
  near-inertial waves in anticyclones. {\em J. Fluid Mech.\/} {\bf 773}, R2.

\bibitem[Eyink(1996)]{Eyink1996}
{\sc Eyink, E.~L.} 1996 Exact results on stationary turbulence in 2{D}:
  consequences of vorticity conservation. {\em Physica D\/} {\bf 91}, 97--142.

\bibitem[Falkovich \& Kritsuk(2017)]{Falkovich2017}
{\sc Falkovich, G. \& Kritsuk, A.~G.} 2017 How vortices and shocks provide for
  a flux loop in two-dimensional compressible turbulence. {\em Phys. Rev.
  Fluids\/} {\bf 2}, 092603(R).

\bibitem[Ferrari \& Wunsch(2009)]{Ferr2009}
{\sc Ferrari, R. \& Wunsch, C.} 2009 Ocean circulation kinetic energy:
  reservoirs, sources, and sinks. {\em Annu. Rev. Fluid Mech.\/} {\bf 31},
  962--971.

\bibitem[Gertz \& Straub(2009)]{Gertz2009}
{\sc Gertz, A. \& Straub, D.~N.} 2009 Near-inertial oscillations and the
  damping of midlatitude gyres: a modeling study. {\em J. Phys. Oceanogr.\/}
  {\bf 39}, 2338--2350.

\bibitem[Jacobson {\em et~al.\/}(2008)Jacobson, Milewski \&
  Tabak]{Jacobson2008}
{\sc Jacobson, T., Milewski, P.~A. \& Tabak, E.~G.} 2008 Mixing closures for
  conservation laws in stratified flows. {\em Stud. Appl. Math.\/} {\bf 121},
  89--116.

\bibitem[Kraichnan(1982)]{Kraichnan1967}
{\sc Kraichnan, R.~H.} 1982 Inertial ranges in two-dimensional turbulence. {\em
  Phys. Fluids\/} {\bf 10}, 1417.

\bibitem[Marino {\em et~al.\/}(2015)Marino, Pouquet \& Rosenberg]{Marino2015}
{\sc Marino, R., Pouquet, A. \& Rosenberg, D.} 2015 Resolving the paradox of
  oceanic large-scale balance and small-scale mixing. {\em Phys. Rev. Lett.\/}
  {\bf 114}, 114504.

\bibitem[Pouquet {\em et~al.\/}(2017)Pouquet, Marino, Mininni \&
  Rosenberg]{Pouquet2017}
{\sc Pouquet, A., Marino, R., Mininni, P.~D. \& Rosenberg, D.} 2017 Dual
  constant-flux energy cascades to both large scales and small scales. {\em
  Phys. Rev. Fluids\/} {\bf 29}, 111108.

\bibitem[Rocha(2018)]{Rocha2018t}
{\sc Rocha, C.~B.} 2018 {\em The turbulent and wavy upper ocean: transition
  from geostrophic flows to internal waves and stimulated generation of
  near-inertial waves\/}. PhD thesis, {U}niversity of {C}alifornia {S}an
  {D}iego.

\bibitem[Rocha {\em et~al.\/}(2018)Rocha, Wagner \& Young]{Rocha2018}
{\sc Rocha, C.~B., Wagner, G.~L. \& Young, W.~R.} 2018 Stimulated generation:
  extraction of energy from balanced flow by near-inertial waves. {\em J. Fluid
  Mech.\/} {\bf 847}, 417--451.

\bibitem[Salmon(1988)]{Salmon1988}
{\sc Salmon, R.} 1988 Hamiltonian fluid mechanics. {\em Ann. Rev. Fluid
  Mech.\/} {\bf 20}, 225--256.

\bibitem[Salmon(1998)]{Salmon1998b}
{\sc Salmon, R.} 1998 {\em Lectures on Geophysical Fluid Dynamics\/}. Oxford
  University Press.

\bibitem[Salmon(2013)]{Salmon2013}
{\sc Salmon, R.} 2013 An alternative view of generalized {L}agrangian mean
  theory. {\em J. Fluid Mech.\/} {\bf 719}, 165--182.

\bibitem[Salmon(2016)]{Salmon2016}
{\sc Salmon, R.} 2016 Variational treatment of inertia–gravity waves
  interacting with a quasi-geostrophic mean flow. {\em J. Fluid Mech.\/} {\bf
  809}, 502--529.

\bibitem[Seshasayanan \& Alexakis(2016)]{Seshasayanan2016}
{\sc Seshasayanan, K. \& Alexakis, A.} 2016 Critical behavior in the inverse to
  forward energy transition in two-dimensional magnetohydrodynamic flow. {\em
  Phys. Rev. E\/} {\bf 93}, 013104.

\bibitem[Seshasayanan {\em et~al.\/}(2014)Seshasayanan, Benavides \&
  Alexakis]{Seshasayanan2014}
{\sc Seshasayanan, K., Benavides, S.~Jose \& Alexakis, A.} 2014 On the edge of
  an inverse cascade. {\em Phys. Rev. E\/} {\bf 90}, 051003(R).

\bibitem[Soward \& Roberts(2010)]{Soward2010}
{\sc Soward, A.~M. \& Roberts, P.~H.} 2010 The hybrid {E}uler--{L}agrange
  procedure using an extension of {M}offatt's method. {\em J. Fluid Mech.\/}
  {\bf 661}, 45--72.

\bibitem[Taylor \& Straub(2016)]{Taylor2016}
{\sc Taylor, S. \& Straub, D.} 2016 Forced near-inertial motion and dissipation
  of low-frequency kinetic energy in a wind-driven channel flow. {\em J. Phys.
  Oceanogr.\/} {\bf 46}, 79--93.

\bibitem[Thomas(2012)]{Thomas2012}
{\sc Thomas, L.~N.} 2012 On the effects of frontogenetic strain on symmetric
  instability and inertia--gravity waves. {\em J. Fluid Mech.\/} {\bf 711},
  620--640.

\bibitem[Vanneste(2013)]{Vanneste2013}
{\sc Vanneste, J.} 2013 Balance and spontaneous generation in geophysical
  flows. {\em Annu. Rev. Fluid Mech.\/} {\bf 45}, 147--172.

\bibitem[Wagner \& Young(2015)]{Wagner2015}
{\sc Wagner, G.~L. \& Young, W.~R.} 2015 Available potential vorticity and
  wave-averaged quasi-geostrophic flow. {\em J. Fluid Mech.\/} {\bf 785},
  401--424.

\bibitem[Wagner \& Young(2016)]{Wagner2016}
{\sc Wagner, G.~L. \& Young, W.~R.} 2016 A three-component model for the
  coupled evolution of near-inertial waves, quasi-geostrophic flow and the
  near-inertial second harmonic. {\em J. Fluid Mech.\/} {\bf 802}, 806--837.

\bibitem[Wunsch \& Ferrari(2004)]{Wuns2004}
{\sc Wunsch, C. \& Ferrari, R.} 2004 Vertical mixing, energy, and the eneral
  circulation of the oceans. {\em Annu. Rev. Fluid Mech.\/} {\bf 36}, 281--314.

\bibitem[Xie \& B\"uhler(2019{\natexlab{{\em a\/}}})]{XieBuhler2019b}
{\sc Xie, J.-H. \& B\"uhler, O.} 2019{\natexlab{{\em a\/}}} Third-order
  structure functions for isotropic turbulence with bidirectional energy
  transfer. {\em J. Fluid. Mech.\/} {\bf 877}, R3.

\bibitem[Xie \& B\"uhler(2019{\natexlab{{\em b\/}}})]{Xie2019}
{\sc Xie, J.-H. \& B\"uhler, O.} 2019{\natexlab{{\em b\/}}} Two-dimensional
  isotropic inertia--gravity wave turbulence. {\em J. Fluid. Mech.\/} {\bf
  872}, 752--783.

\bibitem[Xie \& Vanneste(2015)]{Xie2015}
{\sc Xie, J.-H. \& Vanneste, J.} 2015 A generalised-{L}agrangian-mean model of
  the interactions between near-inertial waves and mean flow. {\em J. Fluid
  Mech.\/} {\bf 744}, 143--169.

\bibitem[Young \& {Ben Jelloul}(1997)]{Young1997}
{\sc Young, W.~R. \& {Ben Jelloul}, M.} 1997 Propagation of near-inertial
  oscillations through a geostrophic flow. {\em J. Mar. Res.\/} {\bf 55}~(4),
  735--766.

\bibitem[Zakharov {\em et~al.\/}(2012)Zakharov, L'vov \&
  Falkovich]{Zakharov2012}
{\sc Zakharov, V.~E., L'vov, V.~S. \& Falkovich, G.} 2012 {\em Kolmogorov
  spectra of turbulence {I}: {W}ave turbulence\/}. Springer Science \& Business
  Media.

\end{thebibliography}
	
\end{document}